 \documentclass[final,5p,times,twocolumn,authoryear]{elsarticle}

\usepackage{amssymb}
\usepackage{lipsum}

\usepackage{lineno,hyperref}
\modulolinenumbers
\usepackage{soul}
\usepackage{color}
\usepackage[usenames,dvipsnames]{xcolor}
\usepackage{multirow}
\usepackage{longtable}
\usepackage{graphicx}
\usepackage{longtable}
\usepackage{tikz}
\usepackage{amsmath}

\usepackage[utf8]{inputenc}
\usepackage{graphicx}
\usepackage{tikz}
\newcommand*\emptycirc[1][1ex]{\tikz\draw (0,0) circle (#1);} 
\newcommand*\halfcirc[1][1ex]{%
  \begin{tikzpicture}
  \draw[fill] (0,0)-- (90:#1) arc (90:270:#1) -- cycle ;
  \draw (0,0) circle (#1);
  \end{tikzpicture}}

 \usepackage{tikz}
\usepackage{longtable}
\usepackage{xurl}
\PassOptionsToPackage{hyphens}{url}\usepackage{hyperref}
\PassOptionsToPackage{hyphens}{url}\usepackage{hyperref}

\def\checkmark{\tikz\fill[scale=0.4](0,.35) -- (.25,0) -- (1,.7) -- (.25,.15) -- cycle;} 
\journal{Journal of \LaTeX\ Templates}

\journal{Latex}

\begin{document}

\begin{frontmatter}



\title{Deep Learning Models for Detecting Malware Attacks}


\author{Pascal Maniriho}
%\ead{P.Maniriho@latrobe.edu.au}
\author{Abdun Naser Mahmood}
\author{Mohammad Jabed Morshed Chowdhury}
\address{Department of Computer Science and Information Technology, La Trobe University, Melbourne, VIC, Australia}

\begin{abstract}
Malware is one of the most common and severe cyber-attack today. Malware infects millions of devices and can perform several malicious activities including mining sensitive data, encrypting data, crippling system performance, and many more. Hence, malware detection is crucial to protect our computers and mobile devices from malware attacks. Deep learning (DL) is one of the emerging and promising technologies for detecting malware. The recent high production of malware variants against desktop and mobile platforms makes DL algorithms powerful approaches for building scalable and advanced malware detection models as they can handle big datasets. This work explores current deep learning technologies for detecting malware attacks on the Windows, Linux, and Android platforms. Specifically, we present different categories of DL algorithms, network optimizers, and regularization methods. Different loss functions, activation functions, and frameworks for implementing DL models are presented. We also present  feature extraction approaches and a review of recent DL-based models for detecting malware attacks on the above platforms. Furthermore, this work presents major research issues on malware detection including future directions to further advance knowledge and research in this field.
\end{abstract}



\begin{keyword}
 Malware detection \sep Deep learning \sep Windows security \sep Android security \sep Linux Security \sep Malware analysis \sep Neural networks
%\MSC[2010] 00-01\sep  99-00
\end{keyword}

\end{frontmatter}

%\tableofcontents



\section{Introduction}
\label{introduction}
\begin{figure*}[h]
  \centering
  \includegraphics[width=0.96\linewidth]{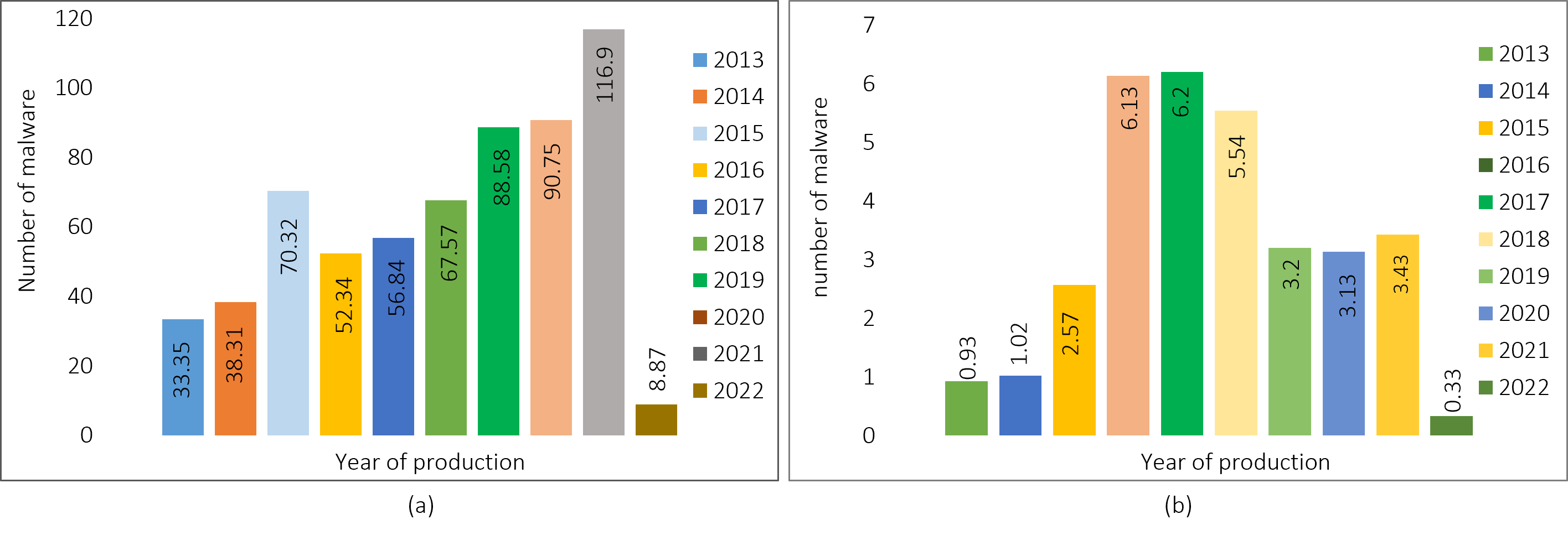}
  \caption{Malware development in (a) Windows from 2013-Feb 2022 \cite{MalwareS17:online} (b) Android from 2013-Feb 2022 \cite{MalwareS17:online}.}
  \label{malware-prod}
\end{figure*}

Recently, malware attacks have become more prevalent and severe cyber threats to today’s Internet security. Malware can steal private sensitive data such as banking details, system login credentials, text messages, and contacts or perform other malicious tasks that can compromise the integrity and availability of data in the infected system \cite{7422770}. Malware attacks have been extremely affecting global economy. As pointed out in \cite{2019202019:online}, about \$6 trillion of the global economy were damaged by cyber-attacks in 2021. The latest security reports have shown that malware continues to be the most prevalent cyber-attack vectors employed by hackers, with ransomware families being the most dangerous attacks that are hard to detect and remove from the compromised systems \cite{Mcafee2021} \cite{checkp2022}. The report in \cite{checkp2022} has also revealed that there is a large production of banking Trojans, backdoors, and potentially forged/piggybacked applications against desktop and mobile devices \cite{checkp2022}. Many Windows, Linux, and Android devices are mostly infected with recent malware attacks \cite{checkp2022}. Figure \ref{malware-prod} (with numbers in million) shows how malware against windows and Android devices have evolved from 2013 - February 2022 \cite{MalwareS17:online}. The number of malware infecting Linux devices also increased by 35\% \cite{Linuxmal57:online}, with new malware families against Linux-based systems appearing in the wild. 

This dramatic growth of malware attacks has created a critical requirement for building advanced malware detection models to protect the existing cyber infrastructures and eliminate the economic impacts associated with these attacks \cite{geluvaraj2019future}. Machine learning (ML) algorithms have been intensively applied to defend cyber infrastructures (internet-connected systems) against malware attacks \cite{gibert2020rise} \cite{muttoo2017android}.  For instance, an ML-based technique was implemented using Support Vector Machine (SVM) to classify malicious Windows files.  K-means and Expectation Maximization algorithms were used by Pai et al. \cite{pai2017clustering} to develop a clustering technique for classifying malware attacks. Accordingly, ML models have proven impeccable performance in solving malware detection problems, and many big companies around the world have been using ML models to detect malware cyber threats against their systems \cite{Microsof19:online} \cite{Secureth26:online} \cite{cybersecurity2017machine}.  Despite the valuable benefits of existing classical ML models, their performance and success highly rely on manually selecting the right features for training and testing the model. This is not an easy task, and it is time-consuming \cite{zhang2021deep}. Additionally, ML models fail to process big datasets \cite{qamar2019mobile}.

Nevertheless, deep learning (DL) models were introduced to address the shortcomings encountered in ML models. DL models are specialized in handling large datasets and performing automatic feature extraction and selection \cite{qamar2019mobile} \cite{ darem2021visualization}. Accordingly, building a malware and classification system with classical ML algorithms requires expertise to design and create feature extraction engines to select good feature representations that the model learns and use the knowledge to classify or detect new input data. Interestingly, this process can be efficiently performed through representation learning, a set of approaches that allows a learning algorithm to receive input of raw data and automatically construct/discover data representations suitable for solving classification or detection tasks \cite{bengio2013representation}.

DL models are categories of representation learning that have many levels of representation designed using simple but non-linear components (also called modules) that transform raw input data from a low representation into a high representation with slightly more abstract features \cite{lecun2015deep} \cite{bengio2013representation}. Hence, with the use of such transformations, DL models can learn very complex patterns from big datasets. There exist many DL-based models presented in the previous works with the aim of detecting malware attacks on various platforms such as Windows, Linux, and Android. These malware detection techniques are mainly grouped into signature-based, behaviour/dynamic-based, and hybrid detection techniques \cite{darem2021visualization} \cite{7814490} \cite{darabian2020detecting}. Recently, DL models for malware detection have advanced to effective techniques based on binary image classification \cite{darem2021visualization}, app's permissions and proprietary Android API package usage \cite{millar2021multi}, and operational code \cite{parildi2021deep}. This work presents a detailed review of current DL technologies which are used to develop malware detection models. Previous state-of-the-art surveys did not adequately cover all DL technologies \cite{wu2021survey} \cite{or2019dynamic} \cite{chakkaravarthy2019survey}. The next section presents various limitations encountered in the previous works and our contributions.

\section{Limitations of previous Surveys and our contributions}
\label{limit-contr}
There exist many surveys on the detection of malware attacks on both desktop and mobile platforms \cite{ucci2019survey} \cite{chakkaravarthy2019survey} \cite{wu2021survey}. Most of them were focused on reviewing machine learning and deep learning-based models, specifically for malware detection and classification. However, none of the previous surveys has adequately elaborated on deep learning technologies for malware detection.  More specifically, as it could be seen from Table \ref{comparison-previous work} which gives a summary of DL technologies and how they were previously covered, various surveys were only focused on malware detection based on classical machine learning algorithms and did not cover current DL technologies. The examples include the surveys/works presented in \cite{black2018survey} \cite{ucci2019survey} \cite{chakkaravarthy2019survey} \cite{8768729} \cite{or2019dynamic} \cite{afianian2019malware}. On the other hand, some of the survey papers have attempted to focus on deep learning, however, they did not cover all deep learning technologies applied in malware detection. Furthermore, they were only limited to elaborating the use of DL models in a single platform such as Android or Windows. For instance, the survey conducted in \cite{wu2021survey} was only focused on DL techniques for malware detection in Android. Nevertheless, topics such as frameworks for building DL models, network optimization, activation, and regularization techniques were not covered. Another example is the survey conducted by Gibert et al. \cite{gibert2020rise} which explored DL techniques for malware detection in the Windows platform, however, their work did not also elaborate on DL technologies such as current academic and industry frameworks for implementing DL models, optimization, and activation techniques.  Moreover Gibert et al.’s work \cite{gibert2020rise} did not also explore malware detection in the Linux and Android platforms. These surveys lack valuable knowledge, the reason why this work is carried out. Overall, the previous surveys highlighted in Table \ref{comparison-previous work} have many limitations which we believe are crucial research gaps in the literature. Therefore, this work aims at presenting a detailed and comprehensive survey of recent DL technologies and their applications for malware detection in Windows, Linux, and Android platforms. These platforms were chosen as they are currently the most platforms targeted and infected by malware cyber-attacks among others. It is worth noting that in Table \ref{comparison-previous work} the symbols \emptycirc[1ex]: means that the topic was not covered in the previous survey, \halfcirc[1ex]: the topic was introduced, however, it was not deeply covered, while \checkmark means that an in-depth coverage on the topic was presented (the topic was deeply covered). The following are our contributions to the literature. The following are our contributions to the literature.

\begin{itemize} 
    \item[$-$] Present emerging deep learning technologies for implementing malware detection models.
    \item[$-$] Review and compile current DL-based techniques for malware detection on desktop and mobile platforms.
    \item[$-$] Present research challenges related to the development of malware detection models based on deep learning algorithms. 
\end{itemize}

\begin{table*}[!]
\centering
\caption{Topics covered in this work against the previous works.} 
\label{comparison-previous work}
%\resizebox{\textwidth}{!}{%
%\begin{tabular}{|l|l|ll|lll|lll|}
\scalebox{0.51}{
\begin{tabular}{|p{26mm}|p{26mm}|p{26mm}|p{23mm}|p{26mm}|p{26mm}|p{75mm}|p{20mm}|p{25mm}|p{20mm}|}
\hline
\multirow{2}{*}{Work presented in} & \multirow{2}{*}{Year} & \multicolumn{2}{c|}{Platform}              & \multicolumn{3}{c|}{Deep Learning Technologies}                                             & \multicolumn{3}{c|}{Current DL-based Detection Techniques}          \\ \cline{3-10} 
                         &                       & \multicolumn{1}{l|}{Desktop OS} & Mobile OS & \multicolumn{1}{l|}{Frameworks for building DL Models} & \multicolumn{1}{l|}{Categories of DL Algorithms} &Optimizers, regularizers, activation and loss functions & \multicolumn{1}{l|}{Windows} & \multicolumn{1}{l|}{Linux} & Android \\ \hline
                         \cite{black2018survey}&   2018                    & \multicolumn{1}{l|}{\checkmark}           &    \emptycirc       & \multicolumn{1}{l|}{ \emptycirc  }           & \multicolumn{1}{l|}{ \emptycirc }           &           \emptycirc                 & \multicolumn{1}{l|}{ \emptycirc }        & \multicolumn{1}{l|}{\emptycirc}      & \emptycirc        \\ \hline
                         \cite{ucci2019survey}&     2019                  & \multicolumn{1}{l|}{\checkmark}           &   \emptycirc         & \multicolumn{1}{l|}{\emptycirc}           & \multicolumn{1}{l|}{\emptycirc}           &      \emptycirc                   & \multicolumn{1}{l|}{\emptycirc}        & \multicolumn{1}{l|}{\emptycirc}      &   \emptycirc       \\ \hline
                         \cite{chakkaravarthy2019survey}&   2019                    & \multicolumn{1}{l|}{\checkmark}           &       \emptycirc      & \multicolumn{1}{l|}{\emptycirc}  & \multicolumn{1}{l|}{\emptycirc }           &  \emptycirc                         & \multicolumn{1}{l|}{\emptycirc}        & \multicolumn{1}{l|}{\emptycirc}      & \emptycirc        \\ \hline
                        \cite{8768729} &   2019                    & \multicolumn{1}{l|}{\emptycirc}           &    \checkmark       & \multicolumn{1}{l|}{\emptycirc}           &  \multicolumn{1}{l|}{\emptycirc}           &        \emptycirc       & \multicolumn{1}{l|}{\emptycirc}        & \multicolumn{1}{l|}{\emptycirc}      &   \emptycirc       \\ \hline
                        \cite{or2019dynamic}&       2019                & \multicolumn{1}{l|}{\checkmark}           &      \checkmark     & \multicolumn{1}{l|}{\emptycirc}           & \multicolumn{1}{l|}{\emptycirc}           &  \emptycirc                       & \multicolumn{1}{l|}{\emptycirc}        & \multicolumn{1}{l|}{\emptycirc}      &    \emptycirc     \\ \hline
                        \cite{afianian2019malware}  &      2019                 & \multicolumn{1}{l|}{\checkmark}           &   \emptycirc        & \multicolumn{1}{l|}{ \emptycirc}           & \multicolumn{1}{l|}{ \emptycirc}           &       \emptycirc                     & \multicolumn{1}{l|}{\emptycirc}        & \multicolumn{1}{l|}{ \emptycirc}      &     \emptycirc       \\ \hline
                        \cite{qamar2019mobile} &     2019                  & \multicolumn{1}{l|}{\emptycirc}   &  \checkmark     & \multicolumn{1}{l|}{\emptycirc}           & \multicolumn{1}{l|}{\halfcirc[1ex]}           &    \emptycirc                   & \multicolumn{1}{l|}{\emptycirc}        & \multicolumn{1}{l|}{\emptycirc}      &  \halfcirc[1ex]   \\ \hline
                        \cite{gibert2020rise} &        2020               & \multicolumn{1}{l|}{\checkmark}           &       \emptycirc   & \multicolumn{1}{l|}{\emptycirc}           & \multicolumn{1}{l|}{\checkmark}   &  \emptycirc      & \multicolumn{1}{l|}{\checkmark}        & \multicolumn{1}{l|}{\emptycirc}      &    \emptycirc     \\ \hline
                        \cite{sahin2020survey} &        2020               & \multicolumn{1}{l|}{\checkmark}           &       \checkmark   & \multicolumn{1}{l|}{\emptycirc}           & \multicolumn{1}{l|}{\halfcirc[1ex]}   &  \emptycirc      & \multicolumn{1}{l|}{\halfcirc[1ex]}        & \multicolumn{1}{l|}{\emptycirc}      &    \halfcirc[1ex]  \\ \hline
                         \cite{qiu2020survey}&   2020                    & \multicolumn{1}{l|}{\emptycirc}           &   \checkmark        & \multicolumn{1}{l|}{\halfcirc[1ex]}           & \multicolumn{1}{l|}{\checkmark}           &  \halfcirc[1ex]                       & \multicolumn{1}{l|}{\emptycirc}        & \multicolumn{1}{l|}{\emptycirc}      &  \checkmark     \\ \hline
                        \cite{9211502} &         2020          & \multicolumn{1}{l|}{\emptycirc}           &   \checkmark        & \multicolumn{1}{l|}{\emptycirc}           & \multicolumn{1}{l|}{\checkmark}           &       \halfcirc[1ex]                  & \multicolumn{1}{l|}{\emptycirc}        & \multicolumn{1}{l|}{\emptycirc}      &    \checkmark    \\ \hline
                        \cite{9118907} &    2020                   & \multicolumn{1}{l|}{\emptycirc}           &   \checkmark        & \multicolumn{1}{l|}{\emptycirc}           & \multicolumn{1}{l|}{\checkmark}           & \emptycirc                         & \multicolumn{1}{l|}{\emptycirc}        & \multicolumn{1}{l|}{\emptycirc}      &   \checkmark      \\ \hline
                        \cite{moussaileb2021survey}&                 2021      & \multicolumn{1}{l|}{\checkmark}           &    \checkmark      & \multicolumn{1}{l|}{\emptycirc}           & \multicolumn{1}{l|}{\emptycirc}           & \emptycirc                        & \multicolumn{1}{l|}{\emptycirc}        & \multicolumn{1}{l|}{\emptycirc}      &   \emptycirc      \\ \hline
                         \cite{sharma2021study}&      2021                 & \multicolumn{1}{l|}{\emptycirc}           &    \checkmark       & \multicolumn{1}{l|}{\emptycirc}           & \multicolumn{1}{l|}{\emptycirc}           &        \emptycirc                 & \multicolumn{1}{l|}{\emptycirc}        & \multicolumn{1}{l|}{\emptycirc}      &     \emptycirc    \\ \hline
                         \cite{singh2021survey}&       2021                 & \multicolumn{1}{l|}{\checkmark}           &   \emptycirc        & \multicolumn{1}{l|}{\emptycirc}           & \multicolumn{1}{l|}{\emptycirc}           &            \emptycirc              & \multicolumn{1}{l|}{\emptycirc}        & \multicolumn{1}{l|}{\emptycirc}      &  \emptycirc  \\ \hline
                         \cite{9591763}&         2021              & \multicolumn{1}{l|}{\checkmark}           &     \emptycirc       & \multicolumn{1}{l|}{\emptycirc}           & \multicolumn{1}{l|}{\emptycirc}           &         \emptycirc               & \multicolumn{1}{l|}{\emptycirc}        & \multicolumn{1}{l|}{\emptycirc}      &      \emptycirc   \\ \hline
                         \cite{abusitta2021malware}&  2021          & \multicolumn{1}{l|}{\checkmark}           &  \emptycirc         & \multicolumn{1}{l|}{\emptycirc}           & \multicolumn{1}{l|}{\halfcirc[1ex]}           &    \emptycirc                     & \multicolumn{1}{l|}{\halfcirc[1ex]}        & \multicolumn{1}{l|}{\emptycirc}      &    \emptycirc     \\ \hline
                         \cite{razgallah2021survey}&          2021             & \multicolumn{1}{l|}{\emptycirc}       &     \checkmark      & \multicolumn{1}{l|}{\emptycirc}           & \multicolumn{1}{l|}{\halfcirc[1ex]}           &           \emptycirc                      & \multicolumn{1}{l|}{\emptycirc}        & \multicolumn{1}{l|}{\emptycirc}      & \halfcirc[1ex]\\ \hline
                         \cite{9585476}&          2021          & \multicolumn{1}{l|}{\emptycirc}           &   \checkmark        & \multicolumn{1}{l|}{\emptycirc}           & \multicolumn{1}{l|}{\halfcirc[1ex]}           &     \emptycirc                    & \multicolumn{1}{l|}{\emptycirc}        & \multicolumn{1}{l|}{\emptycirc}      &  \halfcirc[1ex]       \\ \hline
                         \cite{wu2021survey}&  2021                     & \multicolumn{1}{l|}{\emptycirc}           &     \checkmark      & \multicolumn{1}{l|}{\emptycirc}           & \multicolumn{1}{l|}{\checkmark}           &    \emptycirc                     & \multicolumn{1}{l|}{\emptycirc}        & \multicolumn{1}{l|}{\emptycirc}      &       \checkmark  \\ \hline
                        \cite{sharma2022survey} & 2022                       & \multicolumn{1}{l|}{\emptycirc}           &     \checkmark      & \multicolumn{1}{l|}{\emptycirc}           & \multicolumn{1}{l|}{\halfcirc[1ex]}           &       \emptycirc                  & \multicolumn{1}{l|}{\emptycirc}        & \multicolumn{1}{l|}{ \emptycirc}      &  \halfcirc[1ex]       \\ \hline
                        \cite{galloro2022systematical} & 2022                      & \multicolumn{1}{l|}{\checkmark}           &    \emptycirc       & \multicolumn{1}{l|}{\emptycirc}           & \multicolumn{1}{l|}{\emptycirc}           &             \emptycirc            & \multicolumn{1}{l|}{\emptycirc}        & \multicolumn{1}{l|}{\emptycirc}      &  \emptycirc       \\ \hline
                        The present work & 2022                      & \multicolumn{1}{l|}{\checkmark}           &    \checkmark    & \multicolumn{1}{l|}{\checkmark}           & \multicolumn{1}{l|}{\checkmark}           &    \checkmark          & \multicolumn{1}{l|}{\checkmark}        & \multicolumn{1}{l|}{\checkmark}      &  \checkmark      \\ \hline
\end{tabular}%
}
\end{table*}

\textbf{Organization:} The rest of this article is organized as follows. Section \ref{limit-contr} highlights limitations of the previous works and contributions of this work, Section \ref{background} introduces artificial neural networks and deep learning, Section \ref{dl-tech-types} presents current DL technologies for building malware detection models while, Section \ref{current-dl-tech} discusses feature extraction approaches  and presents a review of current DL-based malware detection models in Windows, Android, and Linux. Section \ref{limit-future-directions} discusses research challenges and feature directions. Section \ref{concl} concludes this work. 

\section{Background}
\label{background}
This Section introduces artificial neural networks and deep learning and also highlights how both techniques are linked.
\subsection{Artificial Neural Networks}
\label{neural-networks}
\begin{figure*}[h]
  \centering
  \includegraphics[width=0.94\linewidth]{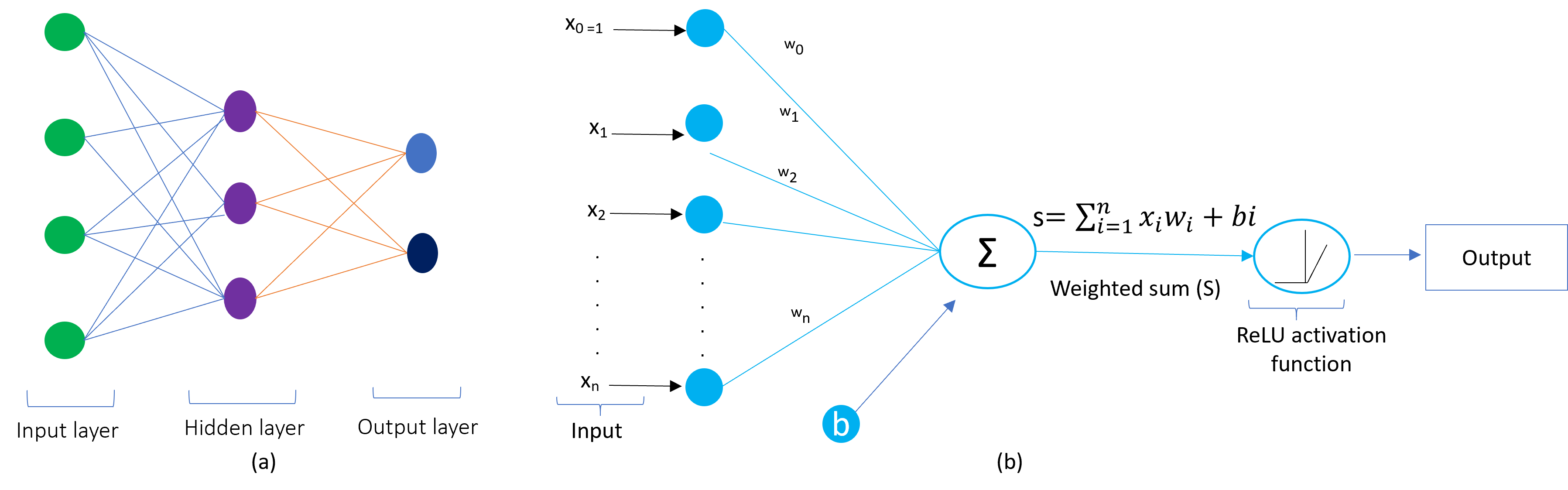}
  \caption{An example of (a) A simple architecture of an artificial neural network (b) Illustration of activation operation in  artificial neural network using ReLU activation function.}
  \label{architecturenn}
\end{figure*}

Artificial neural network (ANN) also known as neural network (NN) is a category of machine learning algorithm that is implemented to mimic how the human brain works. As illustrated in Figure \ref{architecturenn} (a), the basic network architecture of an ANN model has three main layers, namely, the input layer, hidden layer, and output layer. The input neurons receive input data /observations and then pass them to the hidden neurons that perform some operations. The output neurons receive output from the hidden neurons and reveal prediction or classification outcomes. ANNs can be used for performing regression and classification tasks which are achieved through performing various computations on the input data. Given input data $(x)$, the learning process is performed as follows. Small numbers close to zeros are first initialized to be used as weights $(w$) and biases $(b)$ using a uniform or normal distribution \cite{zhang2016gentle}. The second step involves feeding the data/observations to the input layer. Every observation is then forwarded/propagated across the hidden layers which perform various operations to generate the output. This process is commonly known as “forward propagation”. 

\begin{figure*}[h]
  \centering
  \includegraphics[width=0.94\linewidth]{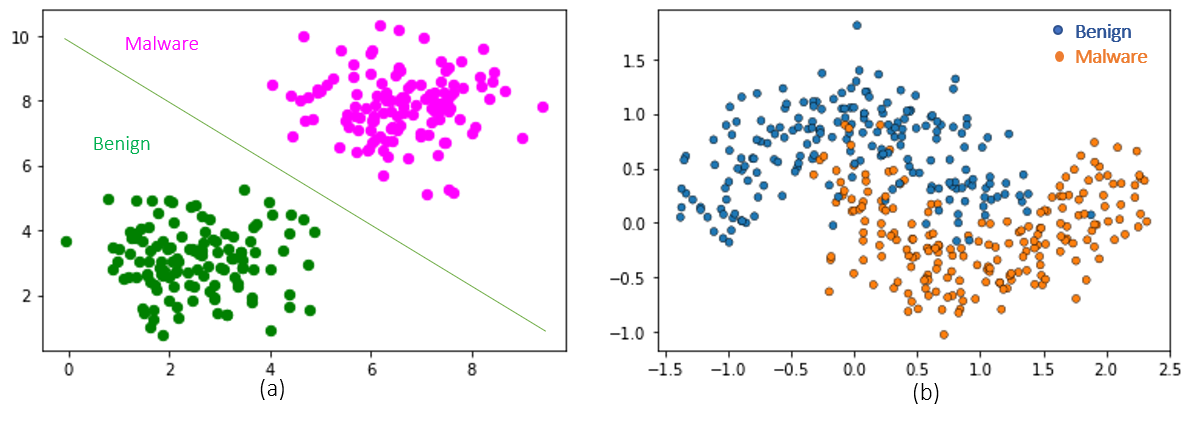}
  \caption{Binary classification (a) Linearly separable classes (b) Non-linearly separable classes.}
  \label{separable}
\end{figure*}

Each output from one layer is activated using an activation function which introduces a non-linearity nature on the network (see Figure \ref{architecturenn} (b)).  This allows the ANN model to identify and learn high-level patterns extracted from the data \cite{apicella2021survey}. Considering a binary classification problem where we have various observations in space with each observation having a label/class, if these observations can be linearly separable into two classes with a straight line or hyperplane in an n-dimensional space, both classes are said to be linearly separable (see Figure \ref{separable} (a) whereas, if a straight line cannot effectively separate both classes, the classification task is a non-linearly separable problem (see Figure \ref{separable} (b)). In a situation like this, a function that introduces a non-linear decision boundary is needed to separate these classes. Thereby, the need for activation functions for neural network models. Sigmoid function and Linear Rectified Unit or simply Rectifier (ReLU) are examples of the most common activation functions \cite{apicella2021survey}. Details on the current activation functions can be found in Section \ref{act-func}. 

It is worth noting that the final output $y$ generated by the last network’s output layer is also fed to an activation function, ensuring that the model produces the expected output. For instance, in a binary classification problem, we expect our ANN model’s output to be the probability of input that belongs to a certain category/class. The Sigmoid function can be used to accomplish this task for a binary classification problem. The predicted value is compared to the actual output (correct class) and the difference between them produces the classification error. To achieve the expected outcome, the neural network model learns by constantly updating or adjusting the weights based on the computed error. This process is performed using a specialized function known as a “loss function or cost function” which minimizes the error, allowing the model to produce accurate prediction \cite{LossandL4:online}. Choosing a loss function highly depends on the type of problem to be solved. After computing the learning error, it is then propagated back to the network’s hidden layers/ to make new adjustments to the weights. This process of adjusting weights based on the computed classification errors is known as “back propagation” and is performed through gradient descent \cite{li2018learning}. Gradient descent is one of the optimization approaches which aims at finding the point with the least/possible minimum error. Back propagation plays an integral part in the performance of ANN classification or predictive models. After adjusting the error to a possible minimum value, the model converges and produces the final classification output. Existing loss functions and network optimization techniques are explored in subsection \ref{optimization-tech}.  Single Layer Perceptron (SLP) \cite{6481037} and Back-Propagation (BP) \cite{Gonzalez2013} are examples of ANN algorithms and were used to detect and classify malware attacks.

\begin{figure*}[!h]
  \centering
  \includegraphics[width=0.94\linewidth]{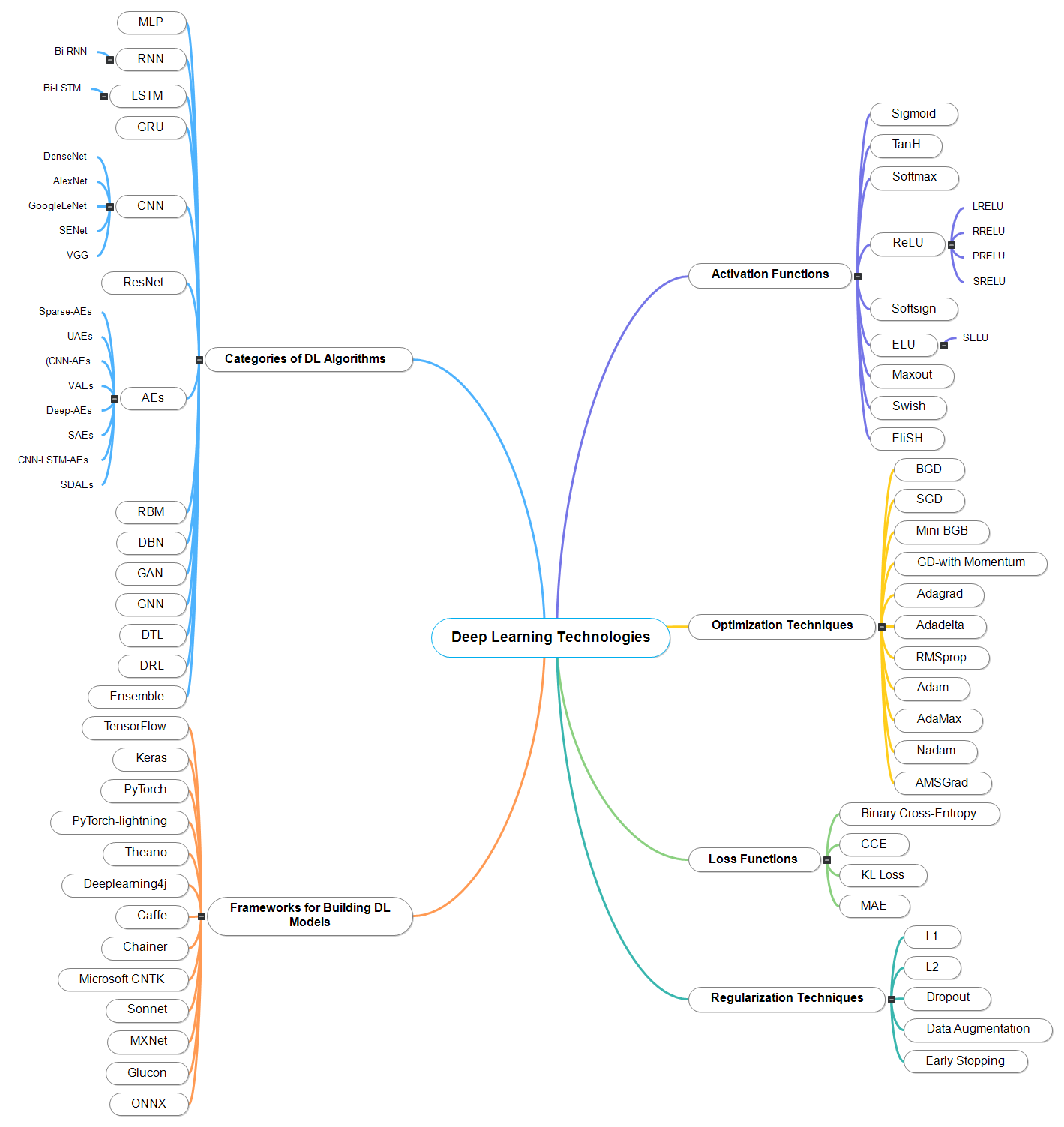}
  \caption{Different deep learning technologies presented in this work.}
  \label{dl-technologies}
\end{figure*}

\subsection{Overview of Deep Learning}
\label{overview dl}
Deep learning is a subset of artificial neural network algorithms that learn from a huge amount of data to perform classification or prediction. Deep learning has become the building block of different key technologies behind computer vision \cite {chai2021deep}, natural language processing \cite{xie2018deep}, autonomous vehicles\cite{9458968}, fraud detection \cite {8566574}, malware detection \cite{mahdavifar2019application} and many more. DL models are characterized by network architectures that contain several hidden layers, the reason why they are often referred to as deep neural networks (DNNs). The term “deep” is usually used to refer to the number of hidden layers in the network.  Deep learning models can perform classification tasks on images, sound, and text data with high classification accuracy. DL network architectures can learn directly from the input datasets and automatically extract features without relying on manual feature extraction, indicating DL’s potential over the traditional neural network algorithms \cite{bengio2013representation}. Moreover, some of the deep learning algorithms can be trained on a computer with a CPU, however, complex deep learning networks require the graphics processing unit (GPU) to accelerate the training process \cite{zhang2021dive}. 

\section{Deep Learning Technologies for Malware Detection}
\label{dl-tech-types}
This section explores current deep learning technologies for malware detection. These include various activation functions, optimization, loss, and regularisation techniques for enhancing the performance of deep learning-based models for malware detection. The types of emerging DL algorithms and frameworks for implementing DL are also presented. Accordingly, Figure \ref{dl-technologies} shows various current DL technologies which are presented in this work.

\subsection{Categories of Deep Learning Algorithms for Malware Detection}
\label{catgory-dl}
In this subsection, we discuss different categories of emerging deep learning algorithms, and further details on their use to detect malware attacks on different platforms are presented in Section \ref{current-dl-tech}.

\subsubsection{Multilayer Perceptron}
\label{mlp}
The multilayer perceptron (MLP) is a subset of feedforward neural networks (FNNs) \cite{8819775} algorithm. It is made up of three types of layers, namely, the input layer, hidden layer, and out layer. Like a feedforward NN, in the MLP neural network architectures, the input data moves in a forward direction from the input layer to the output layer where a backpropagation learning algorithm is used to train the MLP-based model. The MLP models can handle classification problems with non-linearly separable classes and are typically designed to approximate continuous functions \cite{gardner1998artificial}. Nevertheless, MLP models are not specialized in processing sequential data and multidimensional data. In some cases, MLP is combined with other deep learning network architectures such as recurrent neural network (RNN) or convolutional neural network (CNN) \cite{yuan2020byte}  \cite{jeon2020malware}. An MLP algorithm was used to build a stacking ensemble for the Android malware detection framework in the work presented in \cite{9099045}. Using dynamic features such as registry changes, API calls, and network activities, the MLP was employed to implement a binary classification model for classifying in Windows-based systems \cite{singh2021malware}.  .  

\subsubsection{Recurrent Neural Networks}
\label{rnn}
A Recurrent neural network (RNN) is a DL algorithm mainly designed for working with sequential data or time series data. This makes them preferable over traditional feed-forward neural networks that only work for data with features that are independent of each other \cite{alamia2020comparing}. Therefore, if a given dataset has sequential data where one feature value/data point depends on the previous feature values, the network must be modified to allow it to capture dependencies between feature values or data points. To address this problem, RNN algorithms use the concept of “Memory” that enables them to store the information (also called states) about the previous inputs/states to produce the next output of the sequences \cite{bhardwaj2018deep}. Handling sequential data and inputs with varying lengths and storing/memorizing historical information are some of the advantages of RNN \cite{bhardwaj2018deep}. Thereby, RNNs have been used in malware detection. For instance, the work in \cite{jha2020recurrent} has developed a malware detection model that learns from sequences of byte information using RNN and different feature vectorization techniques such as one-hot encoding and random feature vector.  Another RNN-based model was built in \cite{li2021api} using features of API call sequences extracted from portable executable files. Nevertheless, the computation of RNN tends to be slow and classical RNN does not consider future inputs to make decisions. Additionally, the classical RNN algorithm is prone to the vanishing gradient problem, a situation that prevents the model from learning new samples due to the gradients used to compute the weight updates which may get very close to zero \cite{bhardwaj2018deep}. The gradients refer to values used to update the weights of a neural network. The bidirectional recurrent neural networks (BRNN), Gated recurrent unit (GRU), and short-term memory (LSTM) are improved network architectures of classical /regular algorithm.

\subsubsection{Bidirectional Recurrent Neural Networks}
\label{BRNN}
The Bidirectional recurrent neural network (BRNN) was proposed to overcome the shortcomings encountered in regular RNN models.  Different from regular RNN which only operates in the forward mode (where the algorithm learns from the first token of the sequence), the BRNN adds a backward/back-to-from mode that allows the model to also run from the last token of the sequence. To perform such operations more efficiently, the BRNN adds a new hidden layer that allows the information to flow in a backward direction \cite{zhang2021dive}. That is, the BRNN operates by splitting the state neurons of a regular recurrent neural network into two parts, with one part responsible for forward learning (forward states) while the other part is responsible for backward learning (backward states). It is worth noting that the forward states’ outputs are not connected to the inputs of backward states and vice versa.  The same algorithms can be used to train a BRNN model as a unidirectional RNN, given the fact that the states of the neurons in BRNN are not connected \cite{650093}. Pascanu et al. \cite{7178304} proposed a malware detection technique based on bidirectional RNN and their model has achieved good classification results.    
\subsubsection{Long-short Term Memory}
\label{lstm}
Introduced by Hochreiter and Schmidhuber \cite{6795963}, the long-short term memory (LSTM) neural networks are enhanced sequential network architectures developed to solve the problems of vanishing gradient and short-term memory encountered in simple/regular RNNs models. They were mainly designed to work with long sequences of data as they can capture high dependency between elements of a given sequence \cite{6795963}  \cite{lu2019malware}. The LSTM algorithm uses four gates that represent the hidden states of the network to decide about the information to be retained for future predictions \cite{al2019android} \cite{bhardwaj2018deep}. The unidirectional LSTM and bidirectional LSTM (Bi-LSTM) are two main categories of LSTM algorithm \cite{graves2005framewise}. Similar to BRNN, Bi-LSTM processes sequences in forward and backward mode using two non-connected or separate LSTM layers. A multiclass malware classification model based on LSTM was implemented for Windows systems in \cite{andrade2019model}. Song et al. \cite{song2020malicious} proposed a Bi-LSTM model for detecting different attacks such as cross-site request forgery, cross-site scripting, and drive-by download attack, which are the most prevalent malware attacks associated with the JavaScript files.  

\begin{figure*}[h]
  \centering
  \includegraphics[width=0.85\linewidth]{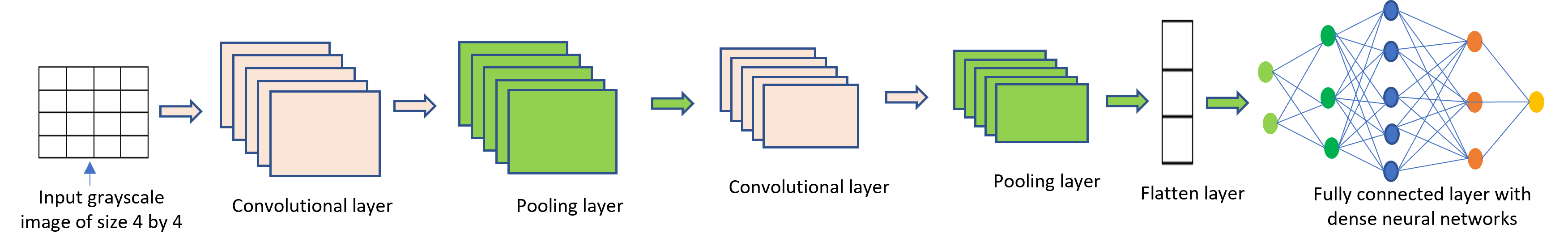}
  \caption{A typical architecture of a CNN model with a grayscale input image of size 4 by 4}
  \label{cnn-arch}
\end{figure*}

\subsubsection{Gated Recurrent Unit}
\label{GRU}
The GRU algorithm is an extended version of LSTM proposed by Kyunghyun et al. in 2014 \cite{cho2014learning}. In contrast to LSTM, the latter uses fewer gated units which reduces the computation overhead but keeps relevant information about the leaned sequences. the GRU network also requires fewer training parameters compared to the LSTM network architecture. The GRU networks also require fewer training parameters compared to the LSTM network architecture and were also introduced to handle the problem of vanishing gradient descent \cite{bhardwaj2018deep} \cite{al2019android}. Al-Thelaya and El-Alfy \cite{al2019android} have used sequences of system call extracted from Android applications in a sandbox environment to implement both unidirectional and bidirectional GRU-based models for Android malware detection. Their experimental results demonstrate better performance over the LSTM-based detection models.

\subsubsection{Convolutional Neural Networks}
\label{cnn}
Like artificial neural networks, convolutional neural networks (CNNs or ConvNets) have neurons with learnable parameters/weights and biases. CNNs were primarily designed for processing images and have been widely applied in computer vision to solve pattern recognition or image classification problems such as object detection and classification \cite{sharma2018analysis}. In recent years, CNN models have been also used in text classification \cite{8912079} and malware detection \cite{vasan2020image}. The network architecture of CNN has three main layers: a convolutional layer, a pooling layer, and a fully connected layer or dense layer (with the same architecture as in ANN or MLP). As depicted in Figure \ref{cnn-arch}, all layers are connected to form a full architecture of CNN. The convolutional layer and pooling layer perform extraction and selection of high compact features/patterns while the dense layer learns from the extracted features to perform classification or prediction.

Malicious and benign binary files were converted into images that were used to build deep CNN-based malware detection models in the work carried out in \cite{vasan2020image}. Kinkead et al. \cite{kinkead2021towards} used the Drebin public benchmark dataset to implement DBN-based malware detection based on sequences of opcodes extracted from Android applications (benign and malware). There exist various forms/architectures of CNN such as DenseNet, AlexNet, GoogLeNet, ResNet, squeeze-and-excitation network (SENet), and VGG, to name a few \cite{huang2017densely} \cite{alzubaidi2021review}, and they have been used for malware detection. For instance, a DenseNet-based model for malware detection was proposed by Hemalatha et al. \cite{hemalatha2021efficient}.

\subsubsection{Deep Residual Networks} 
\label{resnet}
First introduced in 2015 by Ren et al. \cite{he2016deep}, the deep residual network or residual network (ResNet) is one of the advanced CNN Architectures that are widely used to perform image classification tasks. The network architecture of deep ResNet models relies on residual blocks that are built based on the concept of “skip-connections”. The network uses many batch normalizations which allows it to successfully be trained on hundreds of layers without affecting the speed \cite{he2016deep}.  Note that the skip-connections allow the network to skip some layers in-between and are considered the core of residual blocks. Interestingly, deep ResNet has been used for malware detection in recent works. For instance, in Lu et al.’s work \cite{9598532}, raw bytecodes of malware files were converted into RGB Images, and a deep ResNet-18 (network of 18 layers) was used for classifying malware based on images.

\subsubsection{AutoEncoders (AEs)}
\label{Aes}
\begin{figure*}[h]
  \centering
  \includegraphics[width=0.85\linewidth]{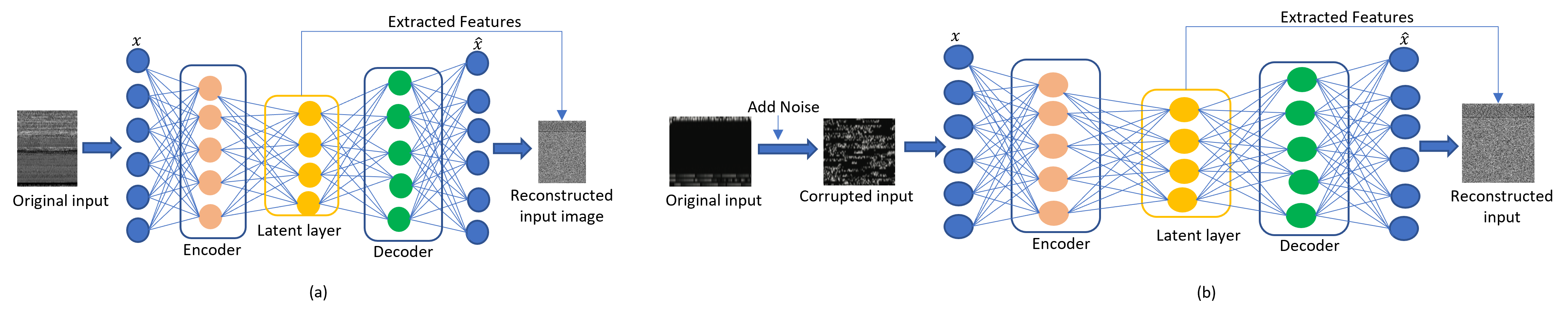}
  \caption{The general network depicting the architecture of (a) generic autoencoders (b) Denoising autoencoders}
  \label{aes-arch}
\end{figure*}

An autoencoder is a type of deep learning algorithm with three-layered network architecture. The network uses an unsupervised learning approach to efficiently learn good feature representation from the input \cite{goodfellow2016deep}. The network architecture of an autoencoder has three main components, namely, an encoder which is a component that processes the input data into an encoded data representation that is smaller than the input (it learns the best features representation), latent/ bottleneck component containing the compressed data representation, and a decoder that allows the network to decompress the compressed data and the original input data. The network architecture has an encoding function and a decoding function. During the learning process, the network is trained to ignore the noise from data while keeping important patterns. The traditional/generic architecture of an autoencoder is depicted in Figure \ref{aes-arch} (a). Nonetheless, the traditional architecture in Figure \ref{aes-arch} (a) has become inefficient in practical applications as the encoding component (hidden layers) tends to copy the same inputs to the output layer \cite {skansi2018introduction}.

 Hence, advanced network architectures such as denoising autoencoders (DAEs) \cite{8474425} were introduced to address this shortcoming. As depicted in Figure \ref{aes-arch} (b), the term “denoising” simply means that the noise is added to the input samples before feeding them to the encoding component. The architecture in Figure \ref{aes-arch} (b) prevents the network from copying the same data to the output, allowing the network to reconstruct good feature representations from the original input data. Moreover, in both Figure \ref{aes-arch} (a) and \ref{aes-arch} (b), the original input samples $x$ are processed by the network to produce output samples $\hat{x}$. Malware detection models based on advanced autoencoders have been presented in recent works and have achieved better performance. Such detection models include the ones based on sparse autoencoders (Sparse-AEs) \cite{d2020malware}, undercomplete autoencoders (UAEs) \cite{9499402} \cite{ray2021new}, convolutional autoencoder(CNN-AEs) \cite{9498570} \cite{wang2019effective}, variational autoencoders (VAEs) \cite{9615643}, deep autoencoders (deep-AEs) \cite{he2018android}, stacked autoencoders (SAEs) \cite{7814490}, CNN-LSTM autoencoders (CNN-LSTM-AEs) \cite{d2021effectiveness}, and stacked denoising autoencoder (SDAEs) \cite{alahmadi2022mpsautodetect}.

\subsubsection{Restricted Boltzmann Machine (RBM)}
\label{rbm}
The restricted Boltzmann machine (RBM in short) is an advanced probabilistic, generative, and unsupervised DL algorithm developed to improve the Boltzmann machine algorithm. RBM learns the joint probability distribution from a given original dataset and uses the knowledge learned to make predictions or inferences on unseen data \cite{liu2021research}. Its network architecture has only two layers, the visible layer/input layer, and the hidden layer that are directly connected. RBM is undirected and is often referred to as an asymmetrical bipartite graph as there is no connection between neurons in the same layer, i.e., one neuron in visible layer cannot be connected to another neuron in the visible layer. Yuan et al. proposed \cite{7399288} DroidDetector, an automated malware detection engine for Android devices based on a stacked ensemble of RBMs models. RBM algorithm is often combined with other deep learning algorithms to improve performance. For instance, Ye et al. \cite{ye2018deepam} presented an intelligent framework for malware detection based on stacked autoencoders, stacked RBM, and association memory algorithms. 

\subsubsection{Deep Belief Networks}
\label{dbns}
A DBN algorithm consists of many restricted Boltzmann machines that are designed by appending a stack of RBM layers. Similar to RBM, neurons/nodes from the same layer cannot communicate However, every layer of the RMB’s network can communicate with both the previous and subsequent layers \cite{saif2018deep}. Except for the first and last layer of the network, each layer of DBN can work as an input to the layer that comes after it or as a hidden layer to the layer that comes after. Recently, DBNs have been widely used in malware detection and classification and have proven to be effective. A DBN-based model for detecting ransomware attacks was proposed in \cite{8556824}. Shi et al. \cite{su2020droiddeep} have used DBN to implement DroidDeep, a new model that characterizes and classifies Android malicious applications with the accuracy of 99.4\%. Saif et al. \cite {saif2018deep} implemented an efficient framework based on deep belief networks for Android-based malware detection systems. This framework achieved an accuracy of 99.1\% and has outperformed some of the existing classical ML-based techniques. 
\subsubsection{Generative Adversarial Networks}
\label{gans}
Generative modeling is a type of unsupervised learning problem that uses generative algorithms to extract important patterns from the input data. A well-designed generative model can generate new plausible (independent) samples or features from real input of original samples \cite{biship2007pattern}. An example of a generative model is Naïve Bayes \cite{xue2008comment} which is often used as a discriminative model (also known as a classification model). On the other hand, deep learning models have been also used as generative models. RBM and DBN are examples of two popular DL generative models. In addition, autoencoders such as stacked denoising autoencoders, deep convolutional generative adversarial networks (DCGANs) \cite{9004932}, and generative adversarial networks (GANs) \cite{8253599}, are examples of the modern DL generative models for malware detection. Moti et al.’s work \cite{moti2019discovering} has generated signatures of unseen malware samples using deep GANs. A detection technique for detecting newly emerging malware attacks (zero-day malware) was implemented using generative adversarial networks and deep autoencoders \cite{kim2018zero}.  

\subsubsection{Graph Neural Networks}
\label{GNNs}
A graph is a data structure made up of two components: nodes (also known as vertices) and edges.  A graph $G$ can be represented by $G= (N, E)$ where $N$ represents a set of nodes while $E$ denotes the edges between them. Graph structures are used to model a set of objects (represented by nodes) and their relationships (represented by edges).  Graph analysis focuses on solving various problems such as node clustering, classifications, and link predictions \cite{zhou2020graph}. Therefore, graph neural networks (GNN) are categories of deep learning models designed to work on the graph domain \cite{zhou2020graph}. Graph convolutional networks (GCN), and graph attention networks (GATNs) are variants of GNNs that have been used to implement malware detection techniques. The study presented by Li et al. \cite{li2021intelligent} has proposed a new intelligent GCN-based model that identifies and classifies malicious files with a detection accuracy of 98.32\%. Another GCN-based detection technique was presented by Pei et al. \cite{pei2020amalnet}.  Hei et al. proposed HawK, a malware detection model that uses GATNs.

\subsubsection{Deep Transfer Learning}
\label{deep-tl}
Deep Learning models perform well when they are trained on large datasets with thousands or even millions of samples before they can effectively make a plausible prediction on unseen samples.  This makes the training process computational expensive, in both training time and required resources.  Consequently, we can take the knowledge learned by one DL model and then transfer it to another DL model to solve different tasks. This learning process is known as deep transfer learning (DTL) \cite{tan2018survey} and has emerged recently. For example, knowledge of a deep neural network model trained to detect anomalies in network traffic can be used in a malware detection system. Several works have explored the use of deep transfer learning models in malware detection. Knowledge from a previously trained deep CNN model was used to build a new image-based deep transfer learning model for detecting malware in the Windows platform \cite{kumar2022dtmic}. Their work demonstrates that using CNN architectures performs well while extracting high-level features from grayscale images of portable executable (PE) files of benign and malware.  The process also saves significant resources and reduces computational time. Chen \cite{chen2018deep} presented a deep transfer learning that detects malware based on static signatures. Using a dataset of 9,339 malware samples from 25 variants, Rezende et al. \cite{8260773} have proposed a DTL model for malware classification developed based on the ResNet-50 pre-trained model. 

\subsubsection{Deep Reinforcement Learning}
\label{drl}
Deep reinforcement learning (DRL) combines deep learning and reinforcement learning (RL) and has been used to solve malware detection problems \cite{sewak2021deep}. Binxiang et al. \cite{9107644} proposed a new DRL-based model that effectively classifies malware attacks in real-time. Further details on the use of DRL for detecting malware attacks can be found in \cite{sewak2021deep}. 

\subsubsection{Ensemble deep learning}
\label{ens-dl}
Ensemble or hybrid DL models are developed using a combination of more than one DL algorithm. For instance, Yan et al.\cite{pei2020amalnet} proposed MalNet, a stacked ensemble based on CNN and LSTM for classifying malicious files. Another DL ensemble model was recently presented by Lin and Chang \cite{lin2021towards} in their work.

\subsection{Optimizers, regularizers, activation and loss functions for DL algorithms}
\label{optimization-tech}
This subsection presents recent activation functions (AFs), network optimizers, loss functions, and regularization techniques that contribute to improving the performance of DL-based models for detecting malware attacks.
\subsubsection{Activation Functions}
\label{act-func}
Activation functions are created to improve the performance of DL models. They are often referred to as “trainable or learnable” activations functions as they have proven to be efficient when improving the performance of DL models \cite{apicella2021survey}. As previously introduced in subsection \ref{neural-networks}, nodes in DL receive $x$ input features that pass through a series of non-linear operations at each neuron’s output until the final prediction is produced. Failure to use the right activation functions can lead to serious problems such as loss of input data or exploding/vanishing the gradients in the model’s network \cite{hayou2019impact}. Thus, below we present recent and commonly used activation functions in DL-based malware detection models.
%Note that the output of activation functions presented in this section is not linear 

\begin{itemize}
\item \textbf{\textit{Sigmoid Function}}–-Mostly denoted by $(sig (x)$ or $\theta (x)$, sigmoid function is a special type of logistic function which is computed using equation \eqref{eq:sigmoid}. It keeps the output of neuron/neural network unit between 0 and 1 during the learning process of a DL model \cite{hertz2018introduction} \cite{szandala2021review} (the function range is 0-1, resulting in an S shape \cite{szandala2021review}). The Sigmoid function is used to stack/dense layers of the network and is among the best choice for binary classification problems. However, it is also employed for non-binary activation. Hard Sigmoid Function, Sigmoid-Weighted Linear Units (SiLU), and  Derivative of Sigmoid-Weighted Linear Units (dSiLU are other variants of Sigmoid activation function \cite{nwankpa2018activation}.
\[
Sig(x) = \frac{1}{1+e^{-x}}    \tag{1} \label{eq:sigmoid}
\]
\item \textbf{\textit{Hyperbolic Tangent Function}}–-Expressed by “tanh or Tanh”, the hyperbolic tangent function (or simply tanh function) \cite{lecun2015deep}, is another alternative that is applied for activation when building DL models. The tanh is quite similar to the Sigmoid function, nevertheless, it has a range of -1 and 1 and it also produces output forming an S shape, which is just an extended sigmoid function. The Tanh function is often preferred over the Sigmoid function, especially for DL models with multi-layer networks \cite{karlik2011performance}. The Hardtanh function (also called Hard Hyperbolic Function) is another form of tanh function employed in DL applications \cite{nwankpa2018activation}. Tanh is computed as shown in equation \eqref{eq:tanh}.
\[
tanh(x)= \frac{e^{x}-e^{-x}}{e^{x}+e^{-x}}    \tag{2} \label{eq:tanh}
\]
\item \textbf{\textit{Softmax}}–-While the Sigmoid activation function is used for binary classification problem, Softmax is another function for multi-class classification tasks \cite{szandala2021review}. Softmax function is presented in equation \eqref{eq:softmx} with $y_i$ representing the softmax, $ z_i $ (input vector), $ e^z_i $ (standard exponential function for input vector), $k$ (number of classes in the multi-class classification problem), and  $ e^z_k $(standard exponential function for output vector). Moreover, Softmax is used for activation in most of the output layers of DL network architectures, however, only when handling multivariate classification problems \cite{simonyan2014very}. 
\[
y_i(z_i) = \frac{e^{z_i}}{ \sum\nolimits_{k=1}^{k}{e^{z_k}} }   \tag{3} \label{eq:softmx}
\]
\item \textbf{\textit{Linear Rectified Unit (ReLU)}}–- This is an advanced activation function that  works by setting all negatives values to zero. Its gradient is computationally very simple compared to gradient descent of Sigmoid and tanh AF \cite{nwankpa2018activation} \cite{szandala2021review} \cite{NairH10}. ReLU performs a simple gradient decent which is 0 or 1 depending on the sign of $x$, resulting in speeding up the training of deep learning models.  ReLU is by far the fastest and the most widely activation function employed in the recent deep neural network models \cite{szandala2021review}. It is mathematically computed using equation in \eqref{eq:relu} where $max(0,1)$ simply means that a value zero (0) will be returned if the input value $(x)$ is negative, otherwise the value is returned. Leaky ReLU (LReLU) \cite{maas2013rectifier}, Parametric Rectified Linear Units (PReLU) \cite{He_2015_ICCV}, Randomized Leaky ReLU (RReLU) \cite{xu2015empirical}, and S-shaped ReLU (SReLU) \cite{jin2016deep} are other variants of ReLU that have been suggested in the literature. 
\[
ReLU(x)= max(0,1)=\left\{\begin{matrix}
 x_{i}, if x_{i} \geqslant 0
\\ 0, if x_{i} <  0             \tag{4} \label{eq:relu}
\end{matrix}\right.
\]
\item \textbf{\textit{Other Activation Functions}}–-There exist other types of network activation functions that are used when implementing DL models. These include functions like Softsign, Softplus Function, Exponential Linear Units (ELUs) with Parametric Exponential Linear Unit (PELU), and Scaled Exponential Linear Units (SELU) being its variants, Maxout Function, Swish Function, and ELiSH \cite{nwankpa2018activation} \cite{apicella2021survey} \cite{szandala2021review} \cite{ramachandran2017searching} and \cite{toth2015phone}. Additionally, it is worth mentioning that ReLU, Logistic Sigmoid, and tanh activations functions may be considered for activations in the hidden layers of deep learning network architectures while Softmax is a good choice for the output layer. Sigmoid also works well for both hidden layer and output layer.

\end{itemize}

\subsubsection{Optimization Techniques}
\label{optimizers}
Deep learning models can learn from experience based on optimization algorithms (or optimizers). Optimization algorithms attempt to minimize/decrease the loss function by computing the gradients \cite{rajendra2021optimization}. Hence, the learning process occurs in several ways with different types of optimizations. In this section, we present optimization algorithms that are used to improve the learning process of deep learning models.  

\begin{itemize}
    \item \textbf{\textit{Batch Gradient Descent}}–-Also known as "vanilla gradient descent", the Batch Gradient Descent (BGD) is a type of gradient descent optimization technique/algorithm that improves the performance of deep learning models/algorithms by utilizing the whole dataset (all samples)  to calculate the cost function’s gradients \cite{ruder2016overview}. That is, BGD uses/considers all samples to take a single step.  The average of the gradients (GDs) is taken for all samples in the training set and then the learning parameters are updated using the computed average of the gradient, resulting in one step of DG in one epoch \cite{ruder2016overview} cite\cite{BatchMin32:online}. Since performing one update or taking one step requires computing the gradients for all samples in the training set, BGD can be very slow and cannot be controlled for big datasets that do not fit well in memory. Additionally, BGD does not also allow updating the model with new samples while training \cite{ruder2016overview}. 
    
    \item \textbf{\textit{Stochastic Gradient Descent}}–-Denoted by SGD, this is another form of gradient descent that computes the gradient of the cost function at each iteration/epoch, making it very fast compared to BGD. Accordingly, the SGD operates as follows. First, it takes a sample from the training set and feed it to the network, compute its gradients, use the computed gradient to updates the network weights, and iteratively, perform these steps for all samples in the training set.  Nevertheless, this will cause the cost to fluctuate over the training samples as one sample is considered at a time, which does not properly minimize the cost function. The cost will decrease with fluctuation after a long run and will never reach the minimum  \cite{BatchMin32:online} \cite{ruder2016overview}.
    
    \item \textbf{\textit{Mini-batch Gradient Descent}}–-It computes the cost function’s gradients using a portion of samples from the training set, i.e., it takes advantage of both BGD and SGD algorithms. An update is performed for each mini-batch of $x$ samples of the training set.  The mini-batch gradient leads to more stable convergence over BGD and SGD and makes the computation of gradients very efficient in deep learning models \cite{ruder2016overview} \cite{BatchMin32:online}.The rest of the optimization algorithms presented below are improvements of the gradient descents (BGD, SGD, and mini-batch gradient) and are widely used by the deep learning community to handle the learning issues mentioned above.  
    
    \item \textbf{\textit{Gradient Decent with Momentum}}–-Also called Momentum, it is an optimization approach that is used to accelerate in the required direction where it operates by fixing the direction to the optimal point \cite{soydaner2020comparison}.
    
    \item \textbf{\textit{Nesterov Accelerated Gradient (NAG)}}–-This optimizer is referred to as SGD with Nesterov momentum or NAG and is also used to improve the direction to the convergence of a DL  algorithm \cite{soydaner2020comparison}. It is a version of the standard momentum that measures the gradient of the loss function slightly ahead in the momentum’s direction instead of the local position. 
    \item \textbf{\textit{Adagrad}}–-The AdaGrad \cite{duchi2011adaptive} is an optimization approach that determines the learning rate based on the learning situation, i.e., since the actual rate is determined from the model’s parameters, the learning rates are adaptive \cite{soydaner2020comparison}. 
    
    \item \textbf{\textit{Adadelta}}–-This optimizer was proposed to enhance the performance of AdaGrad which accumulates the gradients. In contrast, Adadelta uses some windows of fixed size and only tracks the gradients within the windows \cite{soydaner2020comparison} \cite{zeiler2012adadelta}. 
    
    \item \textbf{\textit{RMSprop}}–-This optimization algorithm was also implemented to handle the step size vanishing problem identified in the Adagrad algorithm \cite{hinton2012lecture}. Note that, an exponentially decaying average is used in RMSprop to discard history from the extreme past, helping the algorithm to quickly converge \cite{goodfellow2016deep}.
    
    \item \textbf{\textit{Adam}}–- The Adaptive Movement Estimation (Adam) is an improvement of the SGD optimization algorithm which is used for training deep learning models. It has the best features of the RMSProp and AdaGrad optimizers, making it an advanced and efficient algorithm with the ability to handle sparse gradients \cite{kingma2014adam}. Using Adam, adaptive step size is computed for each learning parameter.
    
    \item \textbf{\textit{AdaMax}}–-This is another extended variant of the Adam optimizer. It works on the basis of the infinity norm such as $L2 norm$ which allows the algorithm to generalize \cite{soydaner2020comparison}. 
    
    \item \textbf{\textit{Nadam}}–-This optimization algorithm was implemented by extending Adam to add the Nesterov momentum (or Nesterov's Accelerated Gradient) algorithm, i.e., the Nesterov momentum was incorporated in Adam. It is an enhanced type of momentum algorithm for deep learning \cite{dozat2016incorporating}. 
    
    \item \textbf{\textit{AMSGrad}}–-This is an improved version of the Adam algorithm. It operates by generalizing to the infinite norm and can be very efficient for some optimization problems in deep neural networks\cite{reddi2019convergence} \cite{Gradient51:online}. 
\end{itemize}

\subsubsection{Network Loss Functions}
\label{loss}
The loss functions (LFs for short) are other important hyperparameters of deep learning algorithms \cite{janocha2017loss}. They allow the DL-based models to minimize the loss or classification error, which enhances the performance on different tasks such as classification. The computed error allows the DL model to update the weights associated with neurons in the network using any of the above network optimizers through backpropagation \cite{janocha2017loss}. The following are the well-known loss functions that are used for computing the error in the DL algorithm for malware classification. 

\begin{itemize}
    \item \textbf{\textit{Binary Cross-Entropy}}–-The Binary Cross-Entropy is one example of Loss Functions (LF) that is heavily used in deep learning models when dealing with classification tasks \cite{afifah2019implementation}. As denoted by $(H_{p}(q))$, the Binary Cross-Entropy is used to compute the classification error when building deep learning models. Accordingly. the equation in \eqref{eq:binloss} is used for computing the binary cross entropy or cross-entropy loss where the parameter $n$ denote the number of samples, $y_{i}$ represents the actual sample’s label, $log(p(y_{i}))$ shows the log probability of a sample/observation to belong to a particular class with $(y=1)$ and $ log(1-p(y_{i})) $  being the probability of an observation to belong to another class $(y=0)$. In \eqref{eq:binloss}, $n$ denotes the number of classes/labels to be predicted which is 2 for binary classification tasks (0 or 1 labels).
\[
H_{p}(q) =-\frac{_{1}^{}}{2}\sum_{i=1}^{n}y_{i}.log(p(y_{i}))+(1-y_{i}).log(1-p(y_{i}))   \tag{5} \label{eq:binloss}
\]
    \item \textbf{\textit{Categorical Cross-Entropy Loss}}–-The Categorical Cross-Entropy Loss (also known as Softmax Loss) is an extension of binary cross-entropy for multiclass classification problems \cite{Understa41:online} \cite{Categori50:online}. The only requirement is that for multiclass classification, only one element will have a non-zero value while other elements in the vector will be zero. This loss function often works in conjunction with the Softmax activation function. The categorical cross-entropy loss function $(CCE)$ computes the loss of a given sample using equation \eqref{cel-cross} in which $\hat{y}$ is the $i^{th}$ scalar in the model’s output, $y$ is the actual target value while $n$ denotes the size of the output vector (number of values in the model’s output. 
\[
CCE=-\sum_{i=1}^{n}y_{i}.log (\hat{y}) \tag{6} \label{cel-cross}
\]
    \item  \textbf{\textit{Other Loss Functions}}–-The Kullback Leibler Divergence Loss (KL Loss) \cite{Understa41:online} and Mean Absolute Error (MAE) \cite{xing2022malware} are other loss functions that are used for malware classifications tasks in DL models.
\end{itemize}

\subsubsection{Regularization Methods for DL Models}
\label{loss}
One of the common issues observed when building deep learning models is overfitting. This situation occurs when a DL model achieves better prediction on training samples but fails to perform well on unseen samples (testing set) \cite{tian2022comprehensive}. Preventing the model to overfit training samples is one way to improve the DL model’s prediction. This can be performed using some regularization methods that help to overcome overfitting, thereby improving the performance on unseen samples (they moderate the learning process). Regularization algorithm helps DL model to generalize well on the unseen sample by making slight changes to the learning model.  L1 and L2, dropout, data augmentation, and early stopping are the most used regularization techniques \cite{tian2022comprehensive} \cite{srivastava2014dropout} \cite{Regulari74:online} when building deep learning models. 

\subsection{DL Frameworks for Implementing Malware Detection Models}
\label{dl frameworks}
The availability of DL frameworks facilitates both academic and industry researchers. with DL frameworks, most of the DL algorithms can be easily built via advanced libraries and application programming interfaces (APIs). Using DL frameworks minimizes the required resources to implement DL models. They greatly minimize the development and deployment time for researchers and security professionals. Many DL frameworks currently offer the possibility to employ Graphics Processing Units (GPUs) accelerators to boost/speed up the learning process of DL models using supported interfaces \cite{Comparis81:online} \cite{Whichdee7:online}. Some of the DL frameworks also allow employing optimized GPU-accelerated libraries like CUDA (cuDNN) \cite{nguyen2019machine} \cite{NVIDIAcu15:online}. CuDNN speeds up (accelerates) most of the DL frameworks such as TensorFlow, PyTorch, and Keras, to name a few \cite{NVIDIAcu15:online}. Table \ref{dl-frameworks} presents current frameworks for building DL models  for malware detection. Table \ref{dl-frameworks} presents current frameworks for building DL models for malware detection.

\begin{table*}[!]
\centering
\caption{Top recent general-purpose frameworks for building deep learning models.}
\label{dl-frameworks}
\resizebox{\textwidth}{!}{%
\begin{tabular}{|l|l|l|l|ll|l|}
%\scalebox{0.70}{
%\begin{tabular}{|p{18mm}|p{38mm}|p{25mm}|p{12mm}|p{12mm}|p{12mm}|p{25mm}|}
%\begin{tabular}{|p{25mm}|p{40mm}|p{30mm}|p{18mm}|p{16mm}|p{16mm}|p{16mm}|}
\hline
\multirow{2}{*}{Framework} & \multirow{2}{*}{Website URL} & \multirow{2}{*}{License} & \multirow{2}{*}{Open-Source} & \multicolumn{2}{c|}{Usage}               & \multirow{2}{*}{Supported Platform} \\ \cline{5-6}
                           &  &       &                & \multicolumn{1}{l|}{Academic} & Industry &                                     \\ \hline
TensorFlow                 &     https://www.tensorflow.org/ &  Apache License 2.0  &     \checkmark & \multicolumn{1}{l|}{ \checkmark}    &  \checkmark  &    Linux, macOS, Windows, Android    \\ \hline
Keras                      &  https://keras.io/           & MIT License   &   \checkmark     & \multicolumn{1}{l|}{\checkmark}         & \checkmark & Cross-platform\\ \hline
PyTorch                    &  https://pytorch.org/     &   Berkeley Software Distribution (BSD)                       &   \checkmark                           & \multicolumn{1}{l|}{\checkmark}         &     \checkmark     &   	Linux, Windows and macOS\\ \hline
pytorch-lightning          & https://www.pytorchlightning.ai/                       &      Berkeley Software Distribution (BSD)               &         \checkmark                     & \multicolumn{1}{l|}{}  &   \checkmark   &  Windows, Linux and macOS                                  \\ \hline
Theano                     &https://theano-pymc.readthedocs.io/                         &  BSD 3-clause license                    &      \checkmark                           & \multicolumn{1}{l|}{\checkmark}         &   \checkmark          &   Linux, macOS and Windows      \\ \hline
Deeplearning4j             &   https://deeplearning4j.konduit.ai/                           &     Apache License 2.0                     & \checkmark                                & \multicolumn{1}{l|}{\checkmark }         &      \checkmark       &                            Cross-platform  \\ \hline
Caffe                      &      https://caffe.berkeleyvision.org/                       &             BSD 2-Clause license               &        \checkmark                         & \multicolumn{1}{l|}{ \checkmark}         &      \checkmark       &     Ubuntu, Red Hat and OS X                               \\ \hline
Chainer                    &     https://chainer.org/                         &     MIT License                     &                \checkmark               & \multicolumn{1}{l|}{\checkmark}         &      \checkmark        &        Cross-Platform                                \\ \hline
Microsoft CNTK             &     	https://www.microsoft.com/en-us/cognitive-toolkit/                         & MIT License      &    \checkmark                          & \multicolumn{1}{l|}{\checkmark}         &     \checkmark     &       Linux  and Windows OS                               \\ \hline
Sonnet                     &    https://sonnet.dev/                          &      Apache License 2.0                    &    \checkmark                          & \multicolumn{1}{l|}{\checkmark}         &   \checkmark       &                macOS and Windows                      \\ \hline
MXNet                      &       https://mxnet.apache.org/                       &                   Apache License 2.0       &        \checkmark                        & \multicolumn{1}{l|}{\checkmark}         &     \checkmark       &     Windows, macOS, Linux                                \\ \hline
Gluon                      &          https://github.com/gluon-api/gluon-api                    &             Apache License 2.0            & \checkmark  & \multicolumn{1}{l|}{\checkmark}         &    \checkmark        &                                  cross-platform     \\ \hline
ONNX                       &          https://onnx.ai/           &               Apache License 2c        &        \checkmark              & \multicolumn{1}{l|}{\checkmark}         &     \checkmark           &                     Windows and Linux                \\ \hline
\end{tabular}%
}
\end{table*}

\section{Detecting  and classifying Malware Attacks}
\label{current-dl-tech}
This section presents feature extraction approaches and recent works on the use of DL algorithms for malware detection. We review current DL-based models/techniques for detecting and classifying malware in the Windows, Linux, and Android platforms.

\subsection{Malware Analysis and Feature Extraction}
\label{progrgam-analyis}
This subsection presents different approaches for analysing malware and features that can be extracted during the analysis of Windows, Android, and Linux files.

\subsubsection{Malware Analysis Approaches}
\label{analysis-approaches}
Discovering the characteristics and objectives of a suspicious file is an important part of malware detection, and the process is known as malware analysis. Static, dynamic, memory and hybrid analysis are types of malware analysis approaches in windows, Linux, and Android \cite{maniriho2021study}.  Static analysis extracts static signatures/patterns from binary files without executing them. In practice, static analysis tends to be simple and very fast, however, it fails to analyse obfuscated malware \cite{biondi2018tutorial}.  In contrast, the dynamic malware analysis allows the malware to execute in an isolated environment while monitoring its behaviours\cite{maniriho2021study}. This makes dynamic analysis resistant to syntactic obfuscation techniques \cite{biondi2018tutorial}. Nevertheless, the dynamic analysis also falls short to track behaviours of very advanced malware such as fileless malware \cite{sihwail2021effective}.  Memory analysis is another approach that can reveal malicious behaviours of fileless malware \cite{sihwail2021effective}. The hybrid malware analysis uses more than one analysis approach and can be efficient over one malware analysis approach \cite{maniriho2021study}.  The features generated after the analysis are used to build malware detection and classification models. 

\begin{table*}[!h]
\caption{Different features extracted in Windows, Android, and Linux files during malware analysis.}
\label{features-extracted}
%\resizebox{\columnwidth}{!}{%
%\begin{tabular}{|l|l|l|l|l|lll|}
\scalebox{0.56}{ 
\begin{tabular}{|p{30mm}|p{38mm}|p{120mm}|p{25mm}|p{25mm}|p{35mm}|p{35mm}|p{35mm}|}

\hline
\multirow{2}{*}{Platform} &
  \multirow{2}{*}{References} &
  \multirow{2}{*}{Extracted Features} &
  \multirow{2}{*}{Static} &
  \multirow{2}{*}{Dynamic} &
  \multicolumn{3}{l|}{Malware/Benign File Type} \\ \cline{6-8} 
                          &  &  &  &  & \multicolumn{1}{l|}{EXE} & \multicolumn{1}{l|}{APK} & ELF \\ \hline
\multirow{13}{*}{Windows} & \cite{li2022novel} \cite{huda2016hybrids}& API call sequences  & \checkmark & \checkmark & \multicolumn{1}{l|}{\checkmark}    & \multicolumn{1}{l|}{}    &     \\ \cline{2-8} 
                          &\cite{huang2021method}  & Images & \checkmark &  & \multicolumn{1}{l|}{\checkmark}    & \multicolumn{1}{l|}{}    &     \\ \cline{2-8} 
                          &\cite{kakisim2022sequential}  & Sequence of opcodes & \checkmark &  & \multicolumn{1}{l|}{\checkmark}    & \multicolumn{1}{l|}{}    &     \\ \cline{2-8} 
                          &\cite{nagano2017static}  & Assembly code and DLL import & \checkmark &  & \multicolumn{1}{l|}{\checkmark}    & \multicolumn{1}{l|}{}    &     \\ \cline{2-8} 
                          & \cite{sun2022leveraging} & Control Flow Graph & \checkmark  &  & \multicolumn{1}{l|}{\checkmark }    & \multicolumn{1}{l|}{}    &     \\ \cline{2-8}
                          & \cite{manavi2022ransomware} & portable executable hear & \checkmark &  &
                          \multicolumn{1}{l|}{\checkmark}    & \multicolumn{1}{l|}{}    &     \\ \cline{2-8} 
                          &\cite{hostiadi2022hybrid}  & Network activities &  & \checkmark  &
                          \multicolumn{1}{l|}{\checkmark }    & \multicolumn{1}{l|}{}    &     \\ \cline{2-8} 
                          & \cite{Singh2020a} & printable strings &  & \checkmark &
                          \multicolumn{1}{l|}{\checkmark}    & \multicolumn{1}{l|}{}    &     \\ \cline{2-8}
                          & \cite{cucchiarelli2021algorithmically} & Malicious DNS &  & \checkmark & \multicolumn{1}{l|}{\checkmark} & \multicolumn{1}{l|}{}    &     \\ \cline{2-8} 
                          & \cite{ghiasi2015dynamic} & Contents of Registers &  &  \checkmark & \multicolumn{1}{l|}{\checkmark }    & \multicolumn{1}{l|}{}    &     \\ \cline{2-8} 
                          & \cite{ijaz2019static} & Registry changes &  & \checkmark  & \multicolumn{1}{l|}{ \checkmark}    & \multicolumn{1}{l|}{}    &     \\ \cline{2-8} 
                          & \cite{jindal2019neurlux} & Loaded DLLs &  &  \checkmark & \multicolumn{1}{l|}{ \checkmark}    & \multicolumn{1}{l|}{}    &     \\ \cline{2-8}
                          &\cite{jindal2019neurlux}  & File System Changes &  &  \checkmark & \multicolumn{1}{l|}{ \checkmark}    & \multicolumn{1}{l|}{}    &     \\ \cline{2-8}
                          
                          & \cite{jindal2019neurlux} & Mutexes &  & \checkmark& \multicolumn{1}{l|}{ \checkmark}    & \multicolumn{1}{l|}{}    &     \\ \cline{2-8}
                          &\cite{santos2013opem} & Running processes and threads &  &  \checkmark & \multicolumn{1}{l|}{ \checkmark}    & \multicolumn{1}{l|}{}    &     \\ \cline{2-8}
                          & \cite{santos2013opem} &  Browsing history&  & \checkmark & \multicolumn{1}{l|}{\checkmark}    & \multicolumn{1}{l|}{}    &     \\ \hline
\multirow{16}{*}{Android} & \cite{ding2016android} & Function calls & \checkmark &  & \multicolumn{1}{l|}{}    & \multicolumn{1}{l|}{\checkmark}    &     \\ \cline{2-8} 
                          & \cite{wei2017machine} & Actions or system events & \checkmark &  & \multicolumn{1}{l|}{}    & \multicolumn{1}{l|}{\checkmark}    &     \\ \cline{2-8} 
                          & \cite{tang2022android} & Operational code (opcode) & \checkmark &  & \multicolumn{1}{l|}{}    & \multicolumn{1}{l|}{\checkmark}    &     \\ \cline{2-8} 
                          & \cite{wang2017detecting} & Strings & \checkmark &  & \multicolumn{1}{l|}{}    & \multicolumn{1}{l|}{\checkmark}    &     \\ \cline{2-8} 
                          & \cite{arshad2018samadroid}& Permissions & \checkmark &  & \multicolumn{1}{l|}{}    & \multicolumn{1}{l|}{\checkmark}    &     \\ \cline{2-8}
                          & \cite{alam2020mining} & control flow (Call graphs) & \checkmark &  & \multicolumn{1}{l|}{}    & \multicolumn{1}{l|}{\checkmark}    &     \\ \cline{2-8} 
                          & \cite{lee2019seqdroid} & Intents & \checkmark &  & \multicolumn{1}{l|}{}    & \multicolumn{1}{l|}{\checkmark}    &     \\ \cline{2-8} 
                          & \cite{tiwari2018android} & API calls &  \checkmark &  & \multicolumn{1}{l|}{}    & \multicolumn{1}{l|}{\checkmark}    &     \\ \cline{2-8}
                          & \cite{yadav2022efficientnet} & Images of APK & \checkmark &  & \multicolumn{1}{l|}{}    & \multicolumn{1}{l|}{\checkmark}    &     \\ \cline{2-8} 
                          & \cite{wang2020deep} &  Infected/malicious URLs& \checkmark &  & \multicolumn{1}{l|}{}    & \multicolumn{1}{l|}{\checkmark}    &     \\ \cline{2-8} 
                          &\cite{wang2019rmvdroid} & app metadata  &  \checkmark  &   & \multicolumn{1}{l|}{}    & \multicolumn{1}{l|}{\checkmark}    &     \\ \cline{2-8} 
                          & \cite{xu2018deeprefiner} & Java byte code & \checkmark &  & \multicolumn{1}{l|}{}    & \multicolumn{1}{l|}{\checkmark}    &     \\ \cline{2-8} 
                          & \cite{zegzhda2017detecting} & Data Flows & \checkmark &  & \multicolumn{1}{l|}{}    & \multicolumn{1}{l|}{\checkmark}    &     \\ \cline{2-8} 
                          & \cite{yan2019understanding} & app's overlay features &  \checkmark &  \checkmark & \multicolumn{1}{l|}{}    & \multicolumn{1}{l|}{ \checkmark}    &     \\ \cline{2-8} 
                          &\cite{yue2017repdroid}&  Runtime User Interface(UI)&  &  \checkmark & \multicolumn{1}{l|}{}    & \multicolumn{1}{l|}{ \checkmark}    &     \\ \cline{2-8} 
                          &\cite{amer2021multi}  & Invoked sequences of API calls &  &  & \multicolumn{1}{l|}{}    & \multicolumn{1}{l|}{}    &     \\ \cline{2-8}
                          & \cite{feng2020two} &Network activities &  &\checkmark  & \multicolumn{1}{l|}{}    & \multicolumn{1}{l|}{\checkmark }    &     \\ \cline{2-8}
                          &\cite{wang2022you} & Runtime permissions  &  & \checkmark  & \multicolumn{1}{l|}{}    & \multicolumn{1}{l|}{\checkmark }    &     \\ \cline{2-8}
                          &\cite{bhandari2018sword}  &Malicious injected code &  & \checkmark & \multicolumn{1}{l|}{}    & \multicolumn{1}{l|}{\checkmark}    &     \\ \cline{2-8}
                          & \cite{massarelli2017android} & Systems/hardware resource consumption((battery, CPU, threads,memory, etc.) &  & \checkmark & \multicolumn{1}{l|}{}    & \multicolumn{1}{l|}{\checkmark}    &     \\ \cline{2-8}
                          & \cite{zhao2017hfa} & communication history(Phone calls,contacted URLs,and SMSs) &  &\checkmark  & \multicolumn{1}{l|}{}    & \multicolumn{1}{l|}{\checkmark}    &     \\ \cline{2-8}
                       & \cite{alhanahnah2020dina} &  Dynamic loaded code &  & \checkmark  & \multicolumn{1}{l|}{}    & \multicolumn{1}{l|}{\checkmark}    &     \\ \hline
\multirow{8}{*}{Linux}    & \cite{xu2021hawkeye} & Assembly instructions & \checkmark &  & \multicolumn{1}{l|}{}    & \multicolumn{1}{l|}{}    &   \checkmark  \\ \cline{2-8} 
                         &\cite{5963012}   & Running processes &  & \checkmark & \multicolumn{1}{l|}{}    & \multicolumn{1}{l|}{}    &  \checkmark   \\ \cline{2-8}
                         & \cite{shalaginov2021novel} & Features from ELF header, subsequent program header table and and  section header table & \checkmark & \checkmark &  \multicolumn{1}{l|}{}    & \multicolumn{1}{l|}{}    &  \checkmark   \\ \cline{2-8} 
                         & \cite{8418602} &Functions and their size  &\checkmark  & & \multicolumn{1}{l|}{}    & \multicolumn{1}{l|}{}    &  \checkmark   \\ \cline{2-8} 
                         & \cite{8418602} & Direct invocation of system calls and PT-loaded segments & \checkmark &  & \multicolumn{1}{l|}{}    & \multicolumn{1}{l|}{}    &  \checkmark   \\ \cline{2-8} 
                         & \cite{8418602} & Overlapping instructions   and number of branch instruction  &\checkmark  &  & \multicolumn{1}{l|}{}    & \multicolumn{1}{l|}{}    &  \checkmark   \\ \cline{2-8}
                         &  \cite{8418602} &  system calls, userspace functions, libraries and loaders&  & \checkmark & \multicolumn{1}{l|}{}    & \multicolumn{1}{l|}{}    & \checkmark    \\ \cline{2-8}
                         &\cite{carrillo2020characterizing} & cyclomatic complexity functions and unique Indicator of compromise (IoC)&  &  \checkmark & \multicolumn{1}{l|}{}    & \multicolumn{1}{l|}{}    &  \checkmark     \\ \cline{2-8}
                         & \cite{monnappa2015automating} & fuzzy hash,network related system calls, and Strings & \checkmark &  & \multicolumn{1}{l|}{}    & \multicolumn{1}{l|}{}    &   \checkmark  \\ \cline{2-8}
                         & \cite{monnappa2015automating} & child process identifier,malicious IP addresses,and network socket created &  & \checkmark & \multicolumn{1}{l|}{}    & \multicolumn{1}{l|}{}    &  \checkmark   \\ \hline
\end{tabular}%
}
\end{table*}

\subsubsection{Feature Extraction}
\label{Features-represent}
This subsection discusses different features that can be extracted to represent malware and benign files in Windows, Android, and Linux platforms. These platforms use different file formats and therefore, features that can be extracted during analysis are different. Table \ref{features-extracted} presents static and dynamic features that can be extracted in Windows executable files (EXE files), Android Package files (package Kit file or APK file), executable and linkable file/ extensible linking files (ELF) in Linux. These files are chosen as they are the most distributed and used files on these platforms.

\subsection{DL-based Malware Detection Models in Windows platform}
\label{win-based-tech}
Given the widespread malware attacks against Windows OS in recent years, building malware detection techniques based on deep learning has been an active topic of interest for both researchers and security professionals. Thus, many DL-based detection models have been proposed. We have collected and compiled recent works on Windows-based malware detection in Table \ref{windows dl}. Table \ref{windows dl} presents various proposed deep learning-based techniques for detecting malware attacks in Windows. It also provides different datasets that have been used for experimental analysis. Below we provide more details on each malware detection technique presented in Table \ref{windows dl}.

Jeon and Moon \cite{jeon2020malware} presented an advanced DL-based malware detection technique using static sequences of opcodes with dynamic RNN and convolutional recurrent neural networks. A convolutional autoencoder was used to transform long sequences of opcodes into short sequences while the recurrent neural network performs malware classification tasks using opcodes features generated by the convolutional autoencoder. Opcode features were statically extracted from windows executable files. Their method achieved the detection accuracy and true positive rate of 96\% and 95\%, respectively. To enhance the performance of malware detection Yuan et al. \cite{yuan2020byte} proposed MDMC, a new model that uses Markov images and CNN to detect malware attacks. The bytes transfer probability matrix was used to convert binary files into Markov images and CNN was used to perform automatic feature engineering and classification. The experiment was carried out using a Microsoft dataset having 10868 malware samples with nine families of malware. The experiment shows better detection performance of MDMC with an accuracy of 99.264\%.

Operational codes (also called Opcodes) extracted from Windows binary executable files were used to build a deep RNN-based model for detecting cryptocurrency malware attacks \cite{yazdinejad2020cryptocurrency}.  Opcodes were extracted from a dataset of 200 benign and 500 cryptocurrency malware samples. Authors have evaluated various structures of LSTM and the model’s performance was validated using a 10-fold cross-validation of the dataset. The experiment shows that the detection model with a three-layered structure achieved an accuracy of 98\%.  Darabian et al. \cite{darabian2020detecting} have used both static features (opcodes) and dynamic features (system calls) extracted from 1500 executable samples of crypto-jacking/crypto-mining malware to implement malware detection models based on CNN, LSTM, and Attention-based LSTM (ATT-LSTM for short). High detection accuracy of 99\% was achieved with system call invocations features while the accuracy rate was 95\% with the static features. The Microsoft dataset was used to develop a multimodal DL-based detection technique that detects malware attacks based on three types of parallel network architectures, namely, a dense network, LSTM, and CNN \cite{9207120}. The implementation was performed using TensorFlow and Keras with Softmax activation function.

% \usepackage{multirow}
\begin{table*}[!h]
\caption{DL-based malware detection models in Windows platform.}
\label{windows dl}
%
}
\end{table*}

Fang et al. \cite{fang2020deepdetectnet} have presented DeepDetectNet, a malware detection model that identifies malware attacks based on static features (imports, entropy, and general file features) extracted from Windows files. This work has also built RLAttackNet, an adversarial DL model that generated samples that bypass the proposed DeepDetectNet. To improve the performance the DeepDetectNet was further trained on the generated adversarial samples which enhanced the value of AUC from 0.989 to 0.996. Khan et al.\cite{9292384} presented malware detection that learns from images of malware and benign program files using deep CNN. Different datasets such as BIG 2015, Malimg, and MaleVis were used in the evaluation, and their method can detect malicious software with a better detection accuracy of 98.65\% obtained using the Malimg dataset. Catak et al.’ work \cite{catak2020deep} has implemented an LSTM-based approach that detects malware with an accuracy of 95\% using a new generated behavioural dataset of API calls extracted from eight families of malware (backdoor, dropper, virus, worm, adware, downloader, spyware, and Trojan). Opcode-level features extracted from obfuscated malware and benign binary files were converted into images to implement a deep CNN-based model that can identify malicious activities of obfuscated malware \cite{darem2021visualization}. The model’s validation demonstrated a detection rate of 99.12\% on the Microsoft BIG-2015 dataset.

Ring et al. \cite {ring2021malware} have extracted process names and accessed file features from audit log events to construct a sequential dataset which was used to develop an LTSM model for malware detection. Various subsects of the dataset were created from the original dataset and each sequence of data was processed using one-hot encoding and embedding representation. Malware detection using sequences of API calls was leveraged in the work presented by Aditya and Girinito \cite{9699248}. Different deep learning network architectures of LSTM and GRU were used to build detection models and their results show that unseen malware can be detected with an accuracy of 96.8\%. An ensemble of dense artificial neural networks and CNN was used to design a stacked malware detection technique that achieves better performance while detecting unknown malware attacks \cite{damavsevivcius2021ensemble}. A multi-tasking DL-based method was implemented using CNN and RGB images of malware and benign files \cite{bensaoud2022deep}. Various optimizers such as Adam, Adagrad, Adadelta RMSprop, and RESprop were tested and the highest detection accuracy (99.97\%) was achieved with Adam. A Deep Transfer learning (DTL)-based malware detection was built using different pre-trained CNN network architectures such as LittleVG, InceptionV3, and ResNet5, to mention a few \cite{alodat2022detection}. The proposed DTL technique was trained and tested on RGB images generated from files of malware and benign. 

Ouahab et al. \cite{ben2022image} employed the GIST descriptor to extract features from malware and clean images provided in the MalImg dataset.  The extracted features were used to train a malware detection model using the multi-layer perceptron (MLP) algorithm with different network architectures. The classification accuracy of 97.6\% was obtained on the selected dataset. Ding et al.\cite{ding2022efficient} collected 20,000 PE malware samples from VirusShare and 20,000 benign PE samples to develop four MalConv detectors for malicious software attacks in the Windows environment. These detectors were implemented with different parameters such as kernel length, stride size, kernel number, and training samples. Ding et al.’s study \cite{ding2022efficient} has also evaluated the effect of adversarial samples on the performance of MalConv detectors. Mallik et al.‘s work \cite{mallik2022conrec} presented ConRec, a convolution recurrent malware detector. Malware samples from Microsoft BIG2015 and MalImg datasets were first converted into grayscale images. Given the high-class imbalance observed between families of malware in both datasets, the data augmentation technique was used to minimize the class bias. VGG16 and Bi-LSTM algorithms were employed to extract features from images of malware samples. The performance was improved through hyperparameter tuning and the results show that their method can effectively identify malware based on their families. 

Mane et al. \cite{mane2022adaptable} employed the MalImg dataset to implement an ensemble-based malware detection method using classical CNN as a feature extractor.  Extracted features were fed to an unsupervised learning module based on K-nearest neighbour (KNN) which categorizes samples in their respective classes. The proposed ensemble method (CNN+KNN) was implemented with less complexity and outperformed the existing detection methods based on VGG16 and ResNet50. The detection accuracy of 99.63\% was achieved by their detection model and the training time was minimized. A Ransomware detection technique based on convolutional neural networks was proposed in \cite{marsh2022ransomware}.  The collected dataset of ransomware consists of six families namely Sage, CTB-Locker, Cerber, CryptoWall, Locky, and TeslaCryp. The detection accuracy of 96\% was obtained on a 10-fold cross-validation with a high true positive rate (TRP) of 95\% and 1\% for false-positive rate (FPR). A framework for detecting malware attacks based on CNN was proposed in \cite{smmarwar2021design}. The byte and assembly features were extracted from the Microsoft BIG2015 dataset. The results show the detection accuracy of 98\% obtained with their detection method. The work in \cite{hemalatha2021efficient} used DenseNet to implement a malware detection method based on the visualization of benign and malware binaries as a two-dimensional image of different dimensions. Microsoft BIG 2015, MalImg, MaleVis, and Malicia datasets were used, and the implementation was performed using the Keras framework.

Benign and malware EXE samples collected from VirusShare were employed to build a VGG16-based detection technique which achieved an accuracy of 94.70\% on hybrid features \cite{huang2021method}. A malware detection model was built using a combination of ResNet50, SENet, and Bi-LSTM with an attention mechanism \cite{jian2021novel}. Microsoft BIG 2015 and NS-Dataset were employed, and samples were converted into three channels of RGB images. Their method detects malware attacks with an accuracy of 97.29\%.  Dynamic sequences of opcodes extracted at runtime were used to train and test different malware detection models (LSTM-RNN, Bi-LSTM, and CNN). The dataset presented in \cite{8027024} was used for analysis while the TensorFlow framework was used in the implementation. A CNN-based technique that detects malware attacks in a metamorphic environment was implemented using TensorFlow and Keras \cite{catak2021data}.  An attention-based deep neural network (ATT-DNNs) for malware detection was presented in \cite{rizvi2022proud} where static features were extracted from EXE files. The detection accuracy of 98.09\% was achieved by their detection model. Local features (sequence of raw codes) and global features (extracted from binary images) were used to train a Generative Adversarial Networks (GANs) model \cite{kim2022obfuscated}. Their method reached an accuracy of 97.47\% using the Microsoft BIG 2015 dataset. In \cite{obaidat2022jadeite} grayscale images were constructed from features extracted from bytecode files using the control flow graph approach and the detection model was built using CNN.  Bytes of malware and benign files were first converted into images to build a GANS and DenseNet-based model for malware detection \cite{tekerek2022novel}. Microsoft BIG 2015 and Dumpware10 datasets were used to evaluate the detection model, and data augmentation was used to handle class imbalance. TensorFlow and Keras were employed in the implementation. Hybrid features were used for developing a CNN-based detection approach using Microsoft BIG2015 and MalImg datasets \cite{li2022malware}.

\subsection{DL-based Malware Detection Models in Android platform}
\label{android-based-tech}
Malware attacks threaten the security of Android devices and have caused significant damage and losses over the past years. Detecting malware in Android devices is still a critical task due to the rapid production of malware variants against Android devices which are increasing rapidly \cite{MalwareS17:online}. Like in Windows-based devices, DL-based models have become one of the security mechanisms against malware attacks on Android devices.Hence, several malware detection models have been proposed. Current DL-based Android models are compiled in Table \ref{andro-works} which includes the proposed detection model and experimental datasets. More importantly, below we provide more details on each malware detection model presented in Table \ref{andro-works}.

Pektas and Acarman \cite{pektacs2020learning} extracted Opcode features from a dataset of 25,000 benign and 24,650 malicious Android apps that were used to train and test a DL-based model. CNN was used to automatically select relevant features and the selected features were passed to an LSTM module to capture more relationships between features extracted by CNN. The final feature set produced by LSTM was passed to a dense layer/fully connected neural network for classification. The grid search approach was employed to identify the best parameters for optimizing the network hyper-parameters while TensorFlow and Keras frameworks were used in the implementation. Their DL method achieved an accuracy of 91.42\% when detecting unknown Android malicious apps.

Ma et al. \cite{ma2020droidetec} proposed Droidect, a Bi-LSTM-based framework that classifies malicious android apps in Android devices.  Behavioural features of API call sequences were extracted from the APK files and were processed using an NLP- semantic localization technique to construct dense vectors which were then fed to the proposed Bi-LSTM model. The experimental evaluation was performed on a dataset of 11982 benign and 9616 malicious files and the detection performance shows an accuracy of 97.22\%. An LSTM-based method was implemented using sequences of opcode extracted from benign and malware applications collected from Play Store and VirusShare, respectively \cite{lakshmanarao2022android}. Text processing techniques were used to process opcodes and Keras was used to build the LSTM malware detector which achieved 96\% detection accuracy. Note that the binary cross-entropy loss function and Adam optimizer were also used to improve the model’s performance. Permissions, API call sequences, intent sequences, and intent filters were employed to implement DL detectors for malware detection in Android using LSTM, GRU, Bi-LSTM, and stacked LSTM/GRU \cite{feng2020seqmobile}. In \cite{darwaish2020rgb} a CNN model was trained on RGB images of benign and malware APK files to build an RGB-based malware detector.  Samples were collected from the AndroZoo dataset, and the experimental results demonstrate high detection accuracy of 99.37\% and a false positive rate of 0.39\%.

\begin{table*}[]
\caption{DL-based malware detection Models in Android platform.}
\label{andro-works}
%
}
\end{table*}

Feng et al. \cite{feng2021android} employed a graph neural network (GNN) and multilayer perceptron (MLP) to implement CGDroid, a lightweight malware detection technique based on static analysis. Call graphs were constructed from each application’s function invocation and intrafunction attributes (security level, required permission, and smali instructions) were further extracted from the node’s attributes of the generated call graph structure using the Word2Vec model. Extracted features were then fed to the GNN module to construct a feature representation passed to the MLP module for classification. Kim et al. \cite{kim2022mapas} proposed MAPAS, a technique that uses call graphs and CNN to classify malicious activities based on behavioural characteristics of malware and benign apps.CNN was used to discover common feature representations from the generated API call graphs while the detection was performed by a lightweight detection module based on the Jaccard similarity coefficient.Their results show that MAPS outdid MaMaDroid \cite{mariconti2016mamadroid} framework with improved accuracy of 91.27\% against 84.99\%, respectively. Mahdavifar et al. \cite{mahdavifar2022effective} presented a new malware detection framework (PLSAE) that relies on DL stacked auto-encoder to detect malicious activities in Android devices. The PLSAE framework uses a semi-supervised learning approach and was trained on hybrid features extracted through static and dynamic analysis. Features were statically and dynamic extracted from apps using CopperDroid, an automated analysis system for Android applications. Extracted feature vectors were passed to a stacked autoencoder (SAE) for feature encoding and the obtained features were fed to a feed-forward Neural Network (FFNN) module for classification. The experimental analysis revealed high detection accuracy of 98.28\% and FPR of 1.16\%, achieved by PLSAE framework \cite{mahdavifar2022effective}. 

Lu et al. \cite{lu2020android} used DBN and GRU algorithms to extract relevant features from raw static and dynamic features. The extracted features were then used to used to train a backpropagation (BP) neural network for malware classification. Their proposed Hybrid DL-based approach has better performance than the conventional ML techniques and can identify malicious apps with obfuscation behaviours. Tensorflow and Keras frameworks were used for implementation. Malware and benign apps were converted into images and CNN was used to develop a detection model which improves the accuracy from 96 \% to 97\% on the dataset collected from Argus Cyber lab \cite{iadarola2021towards}. The static analysis technique was used to extract API patterns from Android apps which were then converted into heterogeneous graphs fed to the graph convolutional neural network (GCN) model for malware detection \cite{gao2021gdroid}. Malicious apps were collected from the ADM dataset and benign apps from Google Play Store while TensorFlow was used to implement the model.

In \cite{almahmoud2021redroiddet}. Permissions, API calls, permission rate, and systems events were used to build the detection model based on RNN Benign and malicious apps were selected from different datasets (CIC-AndMal2017, CIC-InvesAndMal2019, and CIC-MalDroid2020) and the model shows a detection accuracy of 98.58\%. TC-Droid, an automated CNN-based framework for Android malware detection was proposed in \cite{zhang2021deep}. TC-Droid is based on text classification and features such as service, intent, permission, and receiver were extracted from the Android apps collected from Anzhi, Contagio, and Genome. API calls and permissions extracted from benign and malicious apps were used to develop a GRU-based detection model \cite{sabhadiya2019android}. Opcodes, permissions, and GoogleAPIs were extracted from Drebin, AMD, Malgenome, and Intel Security datasets and were used to build a CNN-based malware detector which achieved an accuracy of 99.28\% \cite{millar2021multi}. 

Static features such as intents category, permissions, intents action, and services were used to train an advanced GAN-based detection model implemented using TensorFlow and Keras frameworks \cite{laudanna2021gang}. Bytes and Opcode features were used to implement two DL-based malware detectors (TextCNN and MalConv) using a dataset of malicious apps collected from VirusTotal \cite{yuan2022towards}. Features extracted from Android apps were mined with BERT and XLNET transformer models, and the generated features were employed to implement DL-based methods (MLP, RNN, and TextCNN) for detecting malware attacks \cite{garg2022m2vmapper}. The AMD, CICInvesAndMal2019, and Androzoo datasets were used for experimental analysis.  The work in \ cites{amer2022robust} employed API call sequences and system calls features to detect malicious activities with an LSTM-based detection model where various datasets were used in the experiment. Benign and malicious APK apps were transformed into images which were employed to implement a new malware classification model based on CNN with EfficientNet architecture \cite{yadav2022efficientnet}. Their model was implemented in TensorFlow and Keras while analysis was performed using the R2-D2 dataset. In \cite{kong2022fcscnn} permissions and API calls were extracted from each app and a feature grouping approach was designed to identify relevant features which were employed to build a CNN-based malware classification model. Benign and malicious apps were collected from Google Play Store and VirusTotal, respectively. They have used Tensorflow to build their model which reached an accuracy of 98.07\%. The work presented in \cite{wu2022deepcatra} employed graph neural network (GNN) and Bi-LSTM to develop a multi-view DL approach that identifies Android malware based on API call traces and opcode features. Samples were collected from AndroZoo, Drebin, and DroidAnalytics while TensorFlow and Pytorch were used for implementation. 

Seraj et al. \cite{seraj2022hamdroid} generated a dataset for building malware detection and developed an MLP-based classifier for classifying malware attacks based on permissions features extracted from the apps. Xing et al. \cite{xing2022malware} 
implemented DL-based malware classifiers using different network architectures of DL autoencoders  where all apps were converted into grayscale images. Their proposed model shows an accuracy of 96.22\%. Almomani et al. \cite{9668849} proposed visualization-based malware classification systems for Android which are based on sixteen fine-tuned CNN models such as AlexNet, VGG16, VGG19, MobileNetV2, ResNet50, SqueezeNet, and GoogleNet, to mention a few. Malicious and benign apps were transformed into grayscale and colored visual images and then each image was processed by CNN models for malware classification. The evaluation was conducted on the Leopard Android benchmark dataset with 2486 benign and 14733 malware samples, respectively. Kim et al. \cite{9665792} suggested a multi-streams DL-based approach that classifies malicious apps based on their families. Features represented in string format were first extracted from sections and main files of each benign and malicious app. The extracted features were further used to build a malware classifier using 1D-CNN and 2D-CNN algorithms. The experiments show better performance of 1D-CNN over 2D-CNN.

Ban et al. \cite{ban2022fam} have used the standard architecture of CNN to implement a new framework for Android malware classification. The proposed framework was implemented using TensorFlow and Keras frameworks while features were extracted using Androguard. Their results show a micro F1-score of 0.82 and detection accuracy of 98\%.  The work in \cite{9686531} has presented MemDroid, an LSTM-based detection technique trained on malware samples collected from the Androzoo database. Each app was executed in a sandbox and system call sequences were captured while executing.  All sequences of system calls were used to build the LSTM classifier which distinguishes normal activities from malware activities with a detection accuracy of 99.23\%. In Feng et al.’s work \cite{feng2020performance}  presented MobiTive, a DL-based system that provides real-time detection of malware attacks on Android devices. Different DL architectures such as GUR, LSTM, stacked LSTM/GRU, bidirectional LSTM/GRU, and CNN were used to build the classifier which was further deployed to provide real-time protection for Android devices. 

\subsection{DL-based Malware Detection Models in Linux Platform}
\label{Linux-based-tech} 
The Linux OS has become popular due to its security and scalable features. Linux provides various distributions that support the design of multiple hardware and facilitate the performance of different hardware requirements. These functionalities make the OS the heart of many internet-based desktop devices and create opportunities for cybercriminals. For example, as reported by Palmer \cite{palmer81:online}, the number of malware attacks targeting Linux devices is on the rise in 2022, with cyber-attackers increasingly delivering ransomware and other malware against Linux devices. According to the VMware threat report \cite{VMwareTh56:online}, cryptojacking malware attacks and remote access tools (RATs) are also targeting Linux-based systems. Recently, security researchers have used DL algorithms to detect malware attacks in the Linux environment. 

Xu et al. \cite{xu2021hawkeye} implemented HawkEye, a system that detects malware attacks in Linux based on control flow graph (CFG) and graph neural networks (GNN) and a multilayer perceptron detector (MLP-based classifier). Malware samples were first gathered from AndroZoo and VirusShare repositories while benign samples were collected from executable files and libraries in Ubuntu clean installation. Graph sets were first defined to represent structural information (basic block addresses and assembly instruction/opcodes) from malware and benign executables using a CFG extractor. All generated features were fed to the GNN module to produce graph embedding features which were further fed to the MLP classification module. The evaluation shows a detection accuracy of 96.82\% achieved while detecting malware in a Linux environment.

Landman and Nissim proposed the Deep-Hook framework \cite{landman2021deep} which identifies malware attacks based on captured memory dumps. All dumps were captured during malware and benign execution in a virtual machine (VM), and they were converted into visual images which are used as inputs to CNN for malware classification. The evaluation was conducted on Linux virtual servers and the obtained results show that Deep-Hook can classify malware attacks with a classification accuracy up to 99.9\%. Behavioural features extracted from volatile memory were employed to build a deep neural network-based model that detects attacks in a Linux-based environment \cite{panker2021leveraging}.

 Hwang et al. \cite{hwang2020platform} collected 10,000 Linux malicious binaries files from VirusShare and 10,000 benign files to build a dataset that was used to develop their DNN-based technique for Linux malware detection. During their experiment, 80\% of the dataset (16,000 samples) were used for training the detection model while the remaining portion (20\%) was used for testing. In the work carried out in \cite{shalaginov2021novel}, x86/x64 ELF malware were classified into their families using a DNN-based classification model. Their work has also generated a dataset of 10,574 malware that belongs to 442 different malware families.

\section{Research Challenges and Future Directions}
\label{limit-future-directions}
In this Section, we address the limitations of the previous surveys by only presenting the current main research challenges related to the development of malware detection models using DL technologies. 

\begin{figure*}[h]
  \centering
  \includegraphics[width=0.63\linewidth]{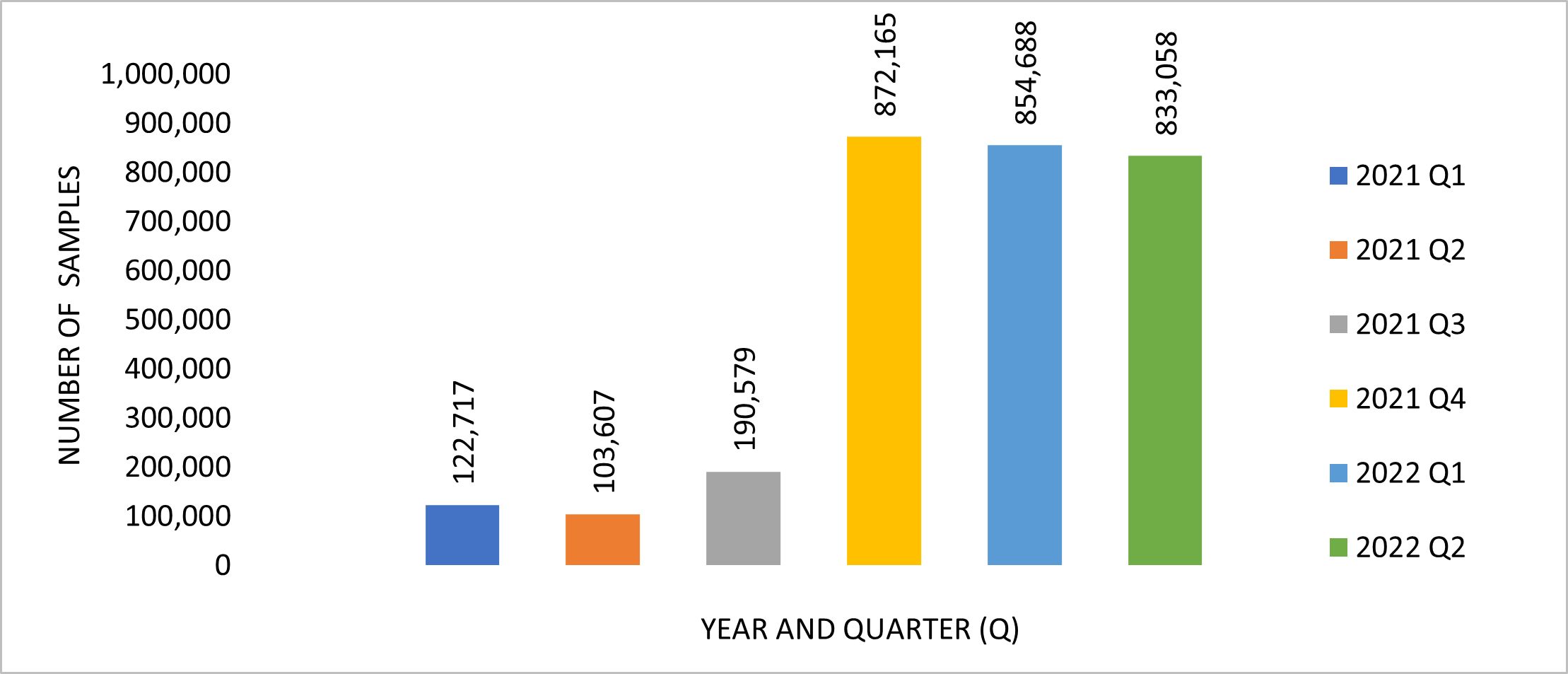}
  \caption{Growth of Linux malware families from 2021 Q1-2022 Q2, data from \cite{Linuxmal29:online}.}
  \label{linux-malware-stat}
\end{figure*}

\subsection{Memorization in Deep Neural Networks}
\label{Memorize-DL}
Existing studies presented in this work show that deep learning models have been widely and used to solve malware classification problems. This remarkable success was mainly achieved due to many parameters used in DL algorithms. To achieve better performance DL network architectures are often trained on large datasets, raising many questions about whether they can go wrong when trained on some types of datasets. Accordingly, the studies presented in \cite{zhang2021understanding} \cite{patel2021memorization} have revealed that the network architectures of DL models can be prone to memorization problems, where the model can end up memorizing the entire dataset resulting in an overfitted model.  Memorization prevents DL models from well generalizing on the training set where the model produces better performance on the training set and poorly performs on the unseen data/test set.  This is a serious problem in the application of DL and can lead to models which are not effective. This problem can be addressed by choosing good regularizes such as data augmentation, weight decay, and dropout and using some DL techniques such as loss functions, which prevent model overfitting \cite{zhang2021understanding} \cite{srivastava2014dropout} \cite{patel2021memorization}.

\subsection{Detection of malware in mobile devices}
\label{maldetect-mov dev}
The number of Android mobile devices hooked to the internment keeps increasing exponentially which arouses developers around the globe to create new applications. For instance, as reported in \cite{Androida33:online}, there are more than 2.65 million Android apps (as of May 2022) on the Google Play Store which are available for download. Unfortunately, the number of cyber-attacks against android devices also continues to grow at the same time \cite{MobileAp26:online}. Despite various malware detection models that have been proposed to secure Android devices, most of the existing DL-based detection models are not real-time and are not suitable for Android mobile devices due to feature complexity/computational cost which are employed for the analysis. In this way, it is ideal to implement new real-time lightweight models for Android mobile devices.

\subsection{Tasks Parallelization in deep learning algorithms}
\label{task-paral}
Despite their valuable application in cybersecurity, training DL models requires a huge amount of time due to many malicious applications which results in big datasets to be processed. On the other hand, scaling network architecture results in a network with complex parameters which also creates time complexity (high execution time while training the model). Fortunately, these problems can be addressed through parallelization mechanisms. Parallelization of tasks in DL models is one of the best approaches for accelerating implementation, i.e., it fastens the algorithm by minimizing the execution time, allowing complex tasks to be processed with less computational resources and execution time \cite{de2012parallelization}. According to the work in \cite{mu2019review}, deep neural networks can be parallelized through data parallelization and model parallelization methods. Therefore, new DL algorithms with high advanced parallelization methods are needed to deliver high-performance DL-based models for malware detection. 

\subsection{Attacks on DL-based Malware Detection Models}
\label{limit3}
Notwithstanding their performance in malware detection, existing works have demonstrated that DL models are vulnerable to adversarial attacks caused by modified input/adversarial samples which are produced by adding imperceptible perturbations to the original input samples where a program is slightly modified to behave differently \cite{suciu2019exploring}. For instance, malicious samples are modified to behave as benign samples while preserving their malicious characteristics. Adversarial samples deceive the performance of DL models, making them a major threat \cite{grosse2017adversarial}. Accordingly, there exist several adversarial attacks introduced against DL models.  Some of the attacks are based on modifying original malware samples by adding non-malicious code (benign code), allowing them to imitate benign programs/applications \cite{kreuk2018deceiving} \cite{suciu2019exploring}. In the work presented in \cite{anderson2018learning} static features from the EXE file header were modified using deep reinforcement learning (DRL) framework to evade a static-based malware detection engine. Semantic NOPs were inserted into the original samples to generate adversarial malware samples that evaded DL-based malware detection techniques implemented using CNN \cite{park2019generation}. Thus, adversarial attacks have created a new major requirement for designing new robust DL-based models that can resist adversarial attacks.

\subsection{Detection of malware in Linux}
\label{maldetec-linux} 
Most of the current studies based on deep learning models were focused on malware detection in both Windows and Android platforms while only a few studies on Dl-based approaches have been presented for detecting malware attacks on the Linux platform. However, sophisticated malware targeting Linux-based systems keeps increasing rapidly \cite{palmer81:online} \cite{VMwareTh56:online} \cite{HowLinux97:online}. As pointed out by Atlas VPN \cite{Linuxmal29:online}, new malicious files targeting Linux has increased in 2022 over 2021 (see Figure \ref{linux-malware-stat}). Researchers at Microsoft also revealed an increase of XorDdos malware attacks targeting Linux devices \cite{Microsof16:online}, arousing security researchers to create new advanced systems for detecting malware in the Linux platform. That is, global organizations need to defend Linux devices against malware attacks as other endpoints devices in the networks. 

\section{Conclusion}
\label{concl}
This work has presented current advances in deep learning for malware detection in Windows, Linux, and Android platforms. Different DL technologies including network optimizers, loss functions, regularization, and activations techniques have been deeply covered. Categories of DL algorithms and frameworks that are essential for the development of DL-based malware detection models were presented. Malware analysis approaches and different features extracted from benign and malware files were discussed. We have reviewed current DL-based models for detecting existing and newly emerging malware variants in the above platforms. The existing datasets for training malware and testing the detection models were also provided. In addition, issues that affect the advancement of malware detention including future directions were discussed. We believe that the work presented in this paper is more relevant and could be one of the steppingstones to advancing the use of deep learning technologies in malware detection in Windows, Linux, and Android platforms. In our future work, we plan to review current advances in the detection of malware attacks on iOS and macOS platforms.

%\section{Summary and conclusions}
%%\label{}
%\lipsum[1-4]

%\section*{Acknowledgements}
%Thanks to ...

%\appendix

%\section{Appendix title 1}

%\section{Appendix title 2}

\bibliographystyle{elsarticle-harv} 
\bibliography{example}



%% For example:

%% \bibitem[Aladro et al.(2015)]{Aladro15} Aladro, R., Martín, S., Riquelme, D., et al. 2015, \aas, 579, A101

%%\end{thebibliography}

\end{document}